\documentclass[12pt,preprint]{aastex}
\usepackage{lscape}

\def\ms{{\rm m\,s^{-1}}}

\def\hbn{{\hfil\break\noindent}}

\begin{document}

\title{Frequency of Hot Jupiters and Very Hot Jupiters from the OGLE-III
Transit Surveys Toward the Galactic Bulge and Carina}

\author{Andrew Gould and Susan Dorsher}
\affil{Dept.\ of Astronomy, Ohio State University,
140 W.\ 18th Ave., Columbus, OH 43210, USA}
\authoremail
{gould,dorsher@astronomy.ohio-state.edu}
\author{B.\ Scott Gaudi}
\affil{Harvard-Smithsonian Center for Astrophysics, 60 Garden Street, Cambridge,
MA 02138, USA}
\authoremail
{sgaudi@cfa.harvard.edu}
\and
\author{Andrzej Udalski}
\affil{Obserwatorium Astronomiczne Uniwersytetu Warszawskiego, 
Aleje Ujazdowskie 4, 00-478 Warszawa, Poland} 
\authoremail
{udalski@astrouw.edu.pl}

\singlespace

\begin{abstract}
We derive the frequencies of hot Jupiters (HJs) with 3--5 day periods
and very hot Jupiters (VHJs) with 1-3 day periods by comparing the
planets actually detected in the OGLE-III survey with those predicted
by our models.  The models are constructed following \citet{gouldandmorgan}
by populating the line of sight with stars drawn from the {\it Hipparcos} 
catalog.  Using these, we demonstrate that the number of stars with
sensitivity to HJs and VHJs
is only 4--16\% of those in the OGLE-III fields satisfying the
spectroscopic-followup limit of $V_{\rm max}<17.5$.  Hence, the frequencies
we derive are much higher than a naive estimate would indicate.
We find that at 90\% confidence the fraction of stars with planets
in the two period ranges is $(1/310)(1^{+1.39}_{-0.59})$
for HJs and $(1/690)(1^{+1.10}_{-0.54})$
for VHJs.  
The HJ rate is statistically
indistinguishable from that found in
radial velocity (RV) studies.  However, we note
that magnitude-limited RV samples are heavily biased toward metal-rich
(hence, planet-bearing) stars, while transit surveys are not, and
therefore we expect that more sensitive transit surveys should find a
deficit of HJs as compared to RV surveys.
The detection of 3 transiting VHJs, all with periods
less than 2 days, is marginally
consistent with the complete absence of such detections in RV surveys.
The planets detected
are consistent with being uniformly distributed between 1.00 and 1.25
Jovian radii, but there are too few in the sample to map this distribution
in detail.  
\end{abstract}

\section{Introduction
\label{sec:intro}}

     More than 150 extrasolar planets have been detected to date, the
majority by radial velocities \citep{butler02,mayor04}, but also by
pulsar timing \citep{wf92}, microlensing \citep{bond04,udalski05}, and
transits \citep{bulge,alpha,udalski02c,systematics,konacki03a,
konacki,konacki-113,tr-113-132,bouchy,tr-111,alonso04}.  These planets and
planetary systems exhibit a wide and unanticipated variety of
characteristics, making the individual detections very
interesting in themselves, to both scientists and the general public.

     However, it is also important to derive from the ensemble of planet
detections (and non-detections), the frequency of planets as functions of
various parameters, such as mass, semi-major axis, and metalicity.
This obviously requires a careful accounting of not only the catalog
of detections, but also the stars for which planets would have been
detected if they had had planets with various specified characteristics.

     In deriving such rates from a given detection technique, the
idiosyncrasies of that technique must be taken into account.
\citet{gaudi00} developed a method for deriving rates for microlensing
searches, which was first implemented by \citet{gaudi02} and later
improved upon by others (\citealt{dong05} and references therein).
Rates have been derived from radial velocity (RV) detections by
\citet{nelson98}, \citet{cumming99}, and \citet{cumming04}, while
\citet{santos05} and \citet{fischervalenti05} have derived relative
frequencies for RV planets as functions of planet and host properties.
The detection efficiencies of various surveys for transiting planets
in clusters have been calculated by \citet{gilliland00}, \citet{mo05},
\citet{weldrake05}, and \citet{burke05}.

     The specific difficulty in deriving rates from transit surveys of
field stars is that the luminosity and radius distributions of the
target stars are not known \citep{gouldandmorgan,brown03}.  
This problem is made substantially worse
by the fact that most transit surveys are directed toward
fields in the Galactic plane where there is a huge background of 
high-luminosity/large-radius sources, which dominate the star counts
but whose planets would be almost completely undetectable.  
These difficulties are
exacerbated by strong selection effects in field surveys such that
the number of expected detections is a strong function of the planet
and host star parameters \citep{scaling,gaudi05,gaudi05a}.

These obstacles must be overcome if the full value of the field
transit surveys is to be realized.  The OGLE survey of
Galactic-plane fields toward Baade's Window and Carina raises
many questions that are difficult to address in the absence
of systematic rate studies.  For example, why does OGLE detect
VHJs even though RV surveys do not?  Why does OGLE seem to detect
so few HJs compared to RV?  Why does OGLE detect comparable numbers
of HJs and VHJs?  Without a well-quantified rate study, it is impossible
to determine if these apparent discrepancies are statistically significant
and, if so, what is their physical origin.

The only
studies to derive quantitative rate information from field transit surveys
\citep{gaudi05,pont} avoided this problem by focusing on the question of
relative frequencies of different classes of planets, rather than determining
absolute frequencies.  This enabled them to address some but not all of these
questions.  Moreover, using relative rates exacerbates uncertainties
due to Poisson fluctuations.  Hence, a more sophisticated treatment is
needed.

     Here we derive absolute frequencies of close-in planets from the
OGLE transit survey.  To determine the distributions of stars in the field as a
function of luminosity and radius, we build on the work of
\citet{gouldandmorgan}, who used the {\it Hipparcos} catalog to model
local field populations.  We generalize their approach to take account
of possible variations in extinction and mean density along various
lines of sight.

In \S\ref{sec:ogledata}, we review the OGLE-III transit data
on which our analysis is based.  In \S\ref{sec:philosophy}, we give
a broad overview of how transiting planets are selected from the data
and how these procedures are simulated in our models.  In 
\S\ref{sec:planetids}, we give a comprehensive summary of the selection
procedures, and in \S\ref{sec:selection} we give an in depth account
of the most important aspects of modeling the resulting selection
effects.  In \S\ref{sec:calculations}, we present our method for
modeling the target populations of the OGLE-III survey, and 
in \S\ref{sec:numprobed} we use our models to evaluate the number
of stars actually probed for planets.  This turns out to be far fewer
than the number in the field satisfying the survey's magnitude limit.
In \S\ref{sec:freqplan}, we evaluate the frequency of planets derived
from the OGLE-III survey and show that the rates are consistent
those derived from RV surveys.  \S\ref{sec:consistency} presents two
Kolmogorov-Smirnov tests that demonstrate that the distributions
of detections as functions of signal-to-noise ratio and $I$-band magnitude
are consistent with the model predictions.  \S\ref{sec:limbdarkening}
gives our prescription for incorporating the effects of limb darkening
and ingress/egress.  These effects, which we demonstrate to be only
a few percent, are actually incorporated in the main analysis, but their
description is deferred to this section for clarity of presentation.
In \S\ref{sec:resonances} and \S\ref{sec:unresolved}, we demonstrate that
effects due to ``resonances'' near integer-day periods and due to unresolved
binaries are small, of order a few percent.  In \S\ref{sec:scaling}, we
investigate how well the scaling relations derived analytically by 
\citet{scaling} for planet senstivity as functions of planet radius and
semimajor axis, apply to the OGLE-III sample.  Finally in 
\S\ref{sec:conclusions}, we summarize our conclusions.

\section{The OGLE Data
\label{sec:ogledata}}

In 2001 and 2002, the OGLE-III project monitored approximately 155,000
disk stars for planetary transits. About 52,000 of these stars
were in three fields toward the Galactic center, and the remaining
103,000 were in three fields toward Carina. The goal of the campaign
was to test the utility of transits as a way to detect short-period
planets. The campaign found both hot Jupiters (HJs), planets with
periods between three and 10 days, whose analogs had previously been
detected in RV surveys, and a new class of planets, the very hot
Jupiters (VHJs), with still shorter periods.

The OGLE-III project observed three slightly overlapping fields toward
the Galactic center between 12 June 2001 and the end of the Galactic
bulge season in October 2001 using the 1.3-m Warsaw telescope at Las
Campanas Observatory, Chile (operated by the Carnegie Institute of
Washington), with a wide field CCD mosaic camera consisting of eight
2048$\times$4096 pixel SITe ST002A detectors. The bulk of the
observations were made in 32 nights spanning the first 45 days, with
single observations of each field taken every few days for the rest of
the season. Altogether, about 800 images per field were taken in
$I$ with 120 second exposures. Additionally, several 150
second $V$-band exposures were taken to provide color
information. However, the zero point of these $V$-band images is
poorly calibrated (0.1 magnitude uncertainty). Because the star field
is quite dense,
no data were taken with seeing worse than $1''\hskip-2pt.8$. The
median seeing was about $1''\hskip-2pt.2$~\citep{bulge}.

Between February and May of 2002, the OGLE-III project observed three
fields toward Carina in the Galactic disk ($l\approx 290^{\circ}$)
using the same telescope and camera as the bulge observations. Over
the span of 95 days, about 1120 $I$-band images per field were
collected. Exposure times were 180 seconds and the median seeing was
about $1''$~\citep{carina}.

OGLE observed an additional six fields in two additional campaigns during
2003 and 2004 \citep{udal04}.
These observations identified 40 additional planetary
transit candidates. However, no RV measurements have been
reported on these candidates, and thus it is not yet known which of these
candidates have planets. We will not consider these two additional OGLE 
campaigns.

\section{Philosophical Approach and Overview
\label{sec:philosophy}}

As a practical matter, the sample is selected in three stages.  First,
OGLE identifies transit candidates derived from its photometric data.  Second,
all of these candidates are examined to identify those for which the
transits can be ascribed to a cause other than a planet.  Finally,
the remaining candidates are observed spectroscopically to determine
if they show RV variations consistent with a planet.  

In order to
extract planet frequencies from a comparison of the model with the
planets detected by this procedure, it is essential that the procedure
itself be well defined, {\it and} that the same
selection criteria are applied to both the data and the model.
This is by no means trivial because, as will become clear, the candidate
sample was not selected by uniform criteria, nor even by a single
group.

Our approach is to systematically review the selection procedures
as reported in the literature and to identify uniform selection
criteria that approximately represent these procedures.  We then
apply these same criteria to the model or, in cases for which
this is impossible, take account of them by making statistical 
corrections to the model.

In order to guide the reader through this fairly involved
exercise, we present in 
Table \ref{tab:vetting} a list of the various selection procedures,
indicating in each case whether they were applied to the data and the model.
In this section, we give a cursory summary of how each selection criterion 
was applied to the data and implemented in the modeling procedure.
Substantially more detail is given for some of these criteria in the
sections that follow.

{\bf Construction of light curves.}  We assume that the conversion of
raw images to light curves is a transparent process that does not
need to be modeled, except that saturated images are systematically
removed.  We identify this saturation limit empirically and impose
it on the model.  In particular, OGLE cleaned the data of systematics
by comparing the behavior of each star with many neighbors, and we
assume that this did nothing to corrupt the transit light curves.

{\bf Automated OGLE Candidate Selection.}  OGLE selected ``pre-candidates''
based on 6 criteria: transit depth $\delta < 0.08$ mag, rms variability
$\sigma<0.015$ mag, period $P$ in the
range 1.052--10.0 days, signal-to-noise parameter $\alpha>9$,
color-magnitude cuts (in the bulge fields only), 
signal detection efficiency $SDE>{\rm max}[3,(4.9 - 0.1\alpha)]$, and
minimum number of transits (3 toward Carina and 2 toward the bulge).
With the exception of the last two, all of these are straightforward
to implement in the model.  However, for three of these, we actually
implemented stronger cuts on the model ($\delta < 0.04$ mag, $P<5$ days,
$\alpha>11$) and excluded any candidates not satisfying these
more restrictive criteria from the data as well.  Whenever a sample
is cut a posteriori, i.e., after it has been searched for candidates,
the results can be biased toward higher detection rates.  We discuss
our reasons for making these more restrictive cuts in some detail
and also argue that any bias is modest compared to the statistical
errors.  We assess the effect of the $SDE$ cut by simulations
and take this effect into account when reporting our derived rates.
We also use simulations to
measure the fraction of otherwise bona fide light curves that have
insufficient transits to be selected by OGLE.

{\bf By-eye OGLE Selection.}  The above automated criteria yielded
several thousand ``pre-candidates'' of which OGLE reported only 137
as ``candidates''.  The remainder were rejected by eye as being
``spurious''.  To model this procedure, one must determine the
fraction of genuine transits that are eliminated in this by-eye step.
We inject transits into real light curves to measure the fraction
for which the transit search algorithm returns the wrong period.
These would almost certainly be lost in the by-eye culling.  For the remainder,
we test directly that the same human (AU) who found the original
transits can also find the injected transits that meet the objective
selection criteria.  Of course, if there were any planets transiting
eclipsing-binary stars, the algorithm would return the binary's period,
and the light curve would be rejected as an eclipsing binary.  We
do not attempt to model this, but such systems are expected to be
rare, in part because the eclipsing binary would often eject the planet.

{\bf Feasibility of Followup.}  For transit candidates to be confirmed
as planets, their small velocity variations must be detected from
high-resolution spectra.  This will be impossible if the stars have
very weak lines, either because they are spinning too fast or are of
too early a type.  As a practical matter, followup would be beyond the
patience of the observers if the star were too faint.  Of course, in
a sample of transit candidates, one expects to find a fair number
of stars that are spinning fast because they are tidally locked to
a close-in massive companion.  These are automatically eliminated from
the data.  However, we make no attempt to account for this in the model,
since the fraction of fast-spinning late-type stars without massive
companions is extremely small.  We model the elimination of early-type
stars by imposing a $(V-I)_0>0.4$ cut in our simulation.  We empirically
determine $V<17.5$ as the effective limit beyond which observers ``refused''
to follow up candidates.

{\bf Non-Eclipsing Binaries.}  Our model assumes that all field stars
are single, whereas more than half of G stars (which we demonstrate
to be the primary target of the OGLE search) are known to be in binaries.
We do not attempt to model this, but show analytically that the impact
on our results is small.

\section{Identification of Transiting Planets
\label{sec:planetids}}

\subsection{Selection Criteria
\label{sec:selcrit}}

We give a brief overview of the selection criteria
and then give details on how they were applied.

1) The OGLE-III survey searched for detections down to a 
signal-to-noise-ratio ($S/N$) parameter $\alpha=9$, but based on
the analysis reported in \S \ref{sec:alphacomplete}
we believe it is complete only to $\alpha_{\rm min}=11$.  In fact,
the completeness is undoubtedly a rising function of $\alpha$ rather
than a strict step function, and our adopted threshold serves as
a proxy for this continuous behavior.  The lowest value for a planet
actually detected by OGLE is $\alpha=11.8$. 

2) OGLE-III searched for periods in the range $1.052<(P/{\rm day})<10$.
However, we further restrict the range to $P<5$ days.  The primary 
reason for this is that, as summarized in \S\ref{sec:summarycand},
there was no systematic followup on 14 OGLE candidates with longer periods,
so no useful scientific conclusions can be derived regarding planets
with periods above 5 days.  In addition, because such planets typically
have very few transits, the modeling approach used in this paper would
not yield accurate results on the survey's sensitivity to them.

3) OGLE-III searched to a transit depth $\delta<0.08$ mag, but we further
restrict this to 0.04 mag.  Our primary reason for this is that,
as summarized in \S\ref{sec:summarycand},
there was no systematic followup on 11 OGLE candidates with deeper transits.
In principle, large planets (or small stars) might yield deeper transits, 
but if so they will be excluded from our sample.  In practice, no planets
have been discovered even near this boundary (despite the survey's
greater sensitivity to them) so this may be a real limit.

4) We require the source color to be $(V-I)_0<0.4$ in order to eliminate
early-F and A stars, which have few lines and so are inaccessible to 
RV followup.

5) We require $V<V_{\rm max}=17.5$ to account for the fact
(discussed in \S\ref{sec:maxdis}) that the observers who
carried out systematic RV followup observations were either
unable or unwilling to monitor stars beyond this limit.

One must be careful when introducing a posteriori selection criteria:
if planets had been discovered just past these boundaries, 
($\alpha<11$, $P> 5$ days, $\delta>0.08$ mag, $(V-I)_0<0.4$,
$V>17.5$) it is unlikely that they would have been excluded
by hypothetical paper writers examining those data.  However,
regarding the period and depth criteria, we really do not have
any choice because the data simply do not exist to conduct studies
beyond our adopted limits.  Regarding $\alpha$, we have done our
best to impartially estimate this boundary.  We also estimate
that the error in our derived rates induced by an error in this
boundary is of order 10\%, which is small compared to the statistical
errors.  The exact color boundary is difficult to estimate precisely,
but there is little doubt that RV followup becomes impossible
substantially beyond our adopted boundary.  The $V$ limit is the 
most difficult to properly estimate, but also has the least impact.
The fact that a substantial number of stars were successfully monitored up 
to $I\sim 16.4$ shows that the effective $V$ limit cannot be too
much brighter.  On the other hand, in \S\ref{sec:goodstars}, we show
that the OGLE-III survey does not have much sensitivity beyond this
value anyway.

Of course, we eliminate all stars for which the transits can be shown
to be due to binaries.  We also eliminate stars with very weak
spectral cross-correlation signatures
regardless of their color.  In principle,
one would like to include these if they meet the other criteria.
However, in practice, it is impossible to determine whether these
stars have a planet or not.  Moreover, given the prior that these
stars have a short-period companion, the most likely explanation
for their indistinct lines is that they are experiencing rapid rotation
induced by tidal locking with a relatively massive companion.

We emphasize that these criteria are not adopted ab initio.  Rather
they are attempts to systematize the criteria that have been applied
in practice by the workers who carried out the survey and followup
observations.

\subsection{Initial OGLE Selection
\label{sec:ogleselect}}

The OGLE-III bulge campaign initially identified 45 distinct sources
that experience shallow-depth transits by cross correlating the
observed light curve with artificial light curves of varying period
containing a 0.015 magnitude transit of duration 0.03 times the
period~\citep{bulge}. After a reanalysis using a search algorithm
developed by~\citet{kovacs}, 13 more candidates for planetary transits
were found in the bulge. The Carina campaign initially identified 62
candidate sources using the same algorithm~\citep{carina}. An
additional 6 bulge sources and 11 Carina sources surfaced after the
correction for small scale systematic effects in both sets of
data~\citep{systematics}. The objects selected by the \citet{kovacs}
algorithm are characterized by two parameters $\alpha$ and $SDE$.

In order to properly simulate the OGLE observations (and so derive
planetary frequencies from them), it is essential to understand both
of these parameters.  The first parameter, $\alpha$, is closely related 
to but not exactly the same as the signal-to-noise ratio $(S/N)$.
Whereas $S/N$ is the ratio of the strength of the signal to the
errors, which are determined from the {\it scatter around the best-fit
transit curve}, $\alpha$ is the ratio of the signal to the {\it scatter around 
a constant flux}.  Since the scatter used to determine $\alpha$ is
inflated by the presence of the transit, $\alpha$ is always
strictly less than $S/N$.  As we derive in \S\ref{sec:maxdis},
\begin{equation}
\alpha = [(S/N)^{-2} + N_{\rm obs}^{-1}]^{-1/2},
\label{eqn:alphaeq}
\end{equation}
where $N_{\rm obs}$ is the total number of observations.  Thus,
for $N_{\rm obs}=1120$, our adopted limit of $\alpha_{\rm min}=11$
corresponds to $(S/N)_{\rm min}=11.6$, a relatively modest difference.  
However,
the difference is accentuated at higher $S/N$ because equation 
(\ref{eqn:alphaeq}) implies that $\alpha$ saturates at $N_{\rm obs}^{1/2}$,
i.e. at $\alpha=33$ in the present example.

The signal detection efficiency ($SDE$) is derived from the
periodogram power spectrum.  It is the ratio of the ``peak excess''
(peak minus the mean) to the standard deviation of this spectrum.
OGLE eliminated candidates that failed to satisfy,
\begin{equation}
SDE > {\rm max}[3,(4.9- 0.1\alpha)].
\label{eqn:sdeeq}
\end{equation}
Correlated errors inject power at many frequencies and thereby increase
the dispersion of the power spectrum and so decrease the $SDE$.  Since,
as we show in \S\ref{sec:sde}, the OGLE errors are correlated,
it will be crucial to determine how the SDE cut affects otherwise
bona fide transit candidates.  

\subsection{{Classes of Spurious Transit Candidates}
\label{sec:spurious}}

Several different physical situations could have been responsible for
the shallow transits seen by the OGLE survey. Some of these physical 
scenarios can be distinguished based on the OGLE observations alone, but
others require additional spectroscopic and/or photometric data. Brown
dwarfs, planets, and low-mass M stars all have about the same radius
and are themselves dim. Thus, they produce comparable flux drops when
they cross a star. Binary systems that are just barely eclipsing will
create a shallow V-shaped dip in the light curve that may be picked up
in a search for shallow transits. To further complicate matters, the
blending of a faint (either dim or distant) binary system with another
star can mimic a shallow transit. The faint binary system experiences
a relatively large change compared to its own flux, but because this
flux is itself small compared to the light added from the brighter
nearby star, the fractional drop in the total flux is small and likely
to be mistaken for a transit.

A variety of techniques can in principle distinguish these backgrounds
from true planetary transits. V-shaped dips can be eliminated by
visual inspection. Some bright-star/eclipsing-binary systems can be
differentiated from planetary systems by measuring the light centroid
shift during the eclipse. If the centroid does not shift, the system
is most likely composed of
a small companion (i.e., a planet, a brown dwarf, or a late
M dwarf) transiting a
star. For an eclipsing binary system along the same line of sight as
another star, the light centroid will most likely shift somewhat
during the eclipse \citep{hoekstra05}. 

\subsection{Culling Candidates Using OGLE Data
\label{sec:oglecull}}

The single most effective technique for distinguishing false positives from
transits is based solely on the OGLE data and eliminates about half
the OGLE-III planetary transit candidates~\citep{drake,sirko}. Stars
with massive companions show variations in their light curves with a
period equal to half the transit period. These variations are caused
by the ellipsoidal shape of a star due to tidal distortion by
its companion. For G0 stars, companions with a mass greater than about
$0.4 M_\odot$ have ellipsoidal variability large enough that it can be
distinguished from noise in the OGLE-III data, allowing these
candidates to be eliminated without the need for followup
measurements~\citep{drake}. Even with these cuts, many cases remain
for which additional photometric and spectroscopic observations are required
to distinguish true planetary transits from various backgrounds.

\subsection{Photometric and Spectroscopic Followup
\label{sec:followup}}

Three groups have used additional photometric data and low resolutions
spectroscopy to further constrain the list of planetary transit
candidates. \citet{dreizler} observed 16 bulge candidates, including
the 13 with the smallest companion radii as predicted by
\citet{bulge}, with the SAAO 1.9m telescope using the Grating
Spectrograph.
They obtained
1800s exposure, 5 \AA\ resolution spectra to classify the primary star
in each system and thus obtain an estimate of its radius. Using this
radius, coupled with the OGLE-III photometric data, they obtained an
estimate of the companion's radius in each system. They found two
objects, OGLE-TR-03 and OGLE-TR-10 with radii of $0.15 R_\odot$, which
they believed were especially likely to be transiting planets.  While
extremely large radii are suggestive that a companion may not be a
planet, the radii of short period planets are still poorly known. A
spectroscopic followup campaign that selected against companions with
large radii could easily miss detecting true planets.

\citet{gallardo} combined the $V$ and $I$ band data from OGLE with
their own $K$ band data taken with the SOFI near-IR array at the 3.5m
ESO New Technology Telescope. 
Exposure times were 3s per image, and the limiting magnitude
was $K\approx 17$. Additionally, they took low resolution spectra with
the Low Dispersion Spectrograph S2 (LDSS2) on the Magellan II 6.5m
Clay telescope on 10 and 11 May 2003 using the $1''$ wide slit with the
high resolution grating, yielding a 6 \AA~resolution. Exposure times
ranged from 90s to 600s. They estimated the effective temperature of the
star in three different distance-dependent ways, such that a distance
that makes the three temperature estimates consistent is likely to be the
true distance of the star. 
One method used the correlation between dereddened
color and effective temperature and the relation between
reddening and distance to extract the effective temperature from
the color.
The second measurement was based upon the main sequence
relation between mass and radius. The third measure (only possible
for some stars) was directly from the low resolution spectrum. Stars
with implausible distance solutions or inconsistent distance solutions
are likely to be giants (OGLE-TR-83, 89, 98, 102, 116, 118, and
119). Once the distance is known, the absolute magnitude of the star
can be calculated, and hence also its radius and the radius of its
transiting companion. \citet{gallardo} found OGLE-TR-105, 109, and 111
to be the most likely to host exoplanets because the radii 
of their transiting companions are less than $0.14\, R_\odot$.

\cite{hh05} obtained low-resolution spectra of OGLE-TR-134 through 137
on 12 and 13 June 2004, using the Boller \& Chivens spectrograph on
the University of Arizona 2.3 m Bok Telescope at Kitt Peak, in order
to spectrally type these candidates.  They were unable
to definitively distinguish between giants and dwarfs, but on the
conservative assumption that the stars were dwarfs, they estimated
the stellar radius from the type and then estimated the planet radius
from the transit depth.  They found best-fit values and $2\,\sigma$ errors
(in units of Jovian radii, $r_J$) of 
$1.33\pm0.11$, $1.43\pm 0.12$, $1.20\pm 0.13$,
and $1.63\pm 0.16$ for the four candidates, respectively.  

\subsection{Radial Velocity Followup
\label{sec:rvfollowup}}

Following the RV confirmation of OGLE-TR-56 by \citet{konacki03a},
which was the first discovery of a planet from transits,
there have been two large campaigns to perform RV
followup measurements of OGLE-III bulge candidates. 
\citet{konacki}
obtained low resolution spectra for 39 of the 59 bulge candidates
after eliminating 20 candidates based on secondary transits, V-shaped
transits, or ellipsoidal variability in the light curves. Brighter
stars were observed with the FAST spectrograph (R=4400) on the 1.5 m
Tillinghast reflector at the Whipple Observatory on Mount Hopkins
in May and June of 2002. The remainder were observed with
the 6.5m Baade telescope Magellan I at Las Campanas Observatory
using the Boller \& Chivens spectrograph (R=2200) between 17
and 21 July 2002. These spectra allowed them to reject eight planetary
transit candidates with early-type primary stars whose transiting
companion must be large (and therefore stellar) 
in order to produce the observed drop in
flux. Additionally, 25 low-depth grazing transits of stars could be
rejected based upon the relatively large RV variations (10 km/s) of
these systems, which could be detected in the low resolution
spectra. Six candidates remained following these
rejections. \citet{konacki} obtained high resolution spectra for five
of these six (OGLE-TR-3, 10, 33, 56, and 58) on the four nights
between 24 and 27 July 2002 using the HIRES instrument (R=65000) at
the Keck I telescope. This group concluded that OGLE-TR-56 has a
Jupiter-size companion, OGLE-TR-58 and OGLE-TR-10 either have
planetary companions or are binary systems blended with brighter
stars, and OGLE-TR-33 is a binary star system blended with a brighter
star. While their own observations of OGLE-TR-3 were inconclusive,
they reanalyzed the data of \citet{dreizler} and concluded that it is
probably a blended system. \citet{konacki-tr-10} later confirmed
OGLE-TR-10 to be an exoplanet transiting its primary star.  In brief,
this campaign confirmed two OGLE-III candidates toward the Galactic
bulge as transiting planets,
OGLE-TR-56 \citep{konacki03a} and OGLE-TR-10 \citep{konacki-tr-10}, while
excluding many others.

\citet{bouchy} performed independent RV followup
observations of the OGLE-III bulge field planetary transit
candidates. In October 2002 they observed three candidates (OGLE-TR-8,
10, and 12) with estimated companion radii less than $1.6\,r_J$
(as reported by \citealt{bulge}) using the UVES spectrograph on the
ESO VLT, achieving a RV precision of $54\,\ms$ on their best
nights and a precision of $93\,\ms$ overall. For each target, they made
8 measurements with exposure times from 20 to 40 minutes. Between May
and June of 2003, they observed 17 candidates (OGLE-TR-5, 6, 7, 10,
12, 17, 18, 19, 33, 34, 35, 48, 49, 55, 56, 58, and 59) with the
FLAMES multi-fiber link to the UVES spectrograph on the VLT, achieving
a precision of about $30\,\ms$. These targets were selected to include
the seven candidates with the smallest estimated radii according to
\citet{bulge} plus a selection of candidates chosen to constrain the
mass-radius relation of low mass stars. \citet{bouchy} 
detected one transiting
planet (OGLE-TR-56), one potential transiting planet that could also
be a blended system (OGLE-TR-10), eight systems with stellar
transiting companions (OGLE-TR-5, 6, 7, 12, 17, 18, 34, and 55), two
grazing eclipsing binaries (OGLE-TR-8 and 35), two triple systems
(OGLE-TR-33 and 59), one false positive (OGLE-TR-58), and three
systems that could not be classified based on their observations
(OGLE-TR-19, 48, and 49). 

\citet{pont} performed RV followup measurements for 42
candidates in the Carina field using the FLAMES facility with the UVES
spectrograph on the VLT during eight half-nights 13 to 21 March
2004. This campaign yielded three confirmations of OGLE-III candidates
as transiting planets, OGLE-TR-113 and OGLE-TR-132 \citep{tr-113-132} and
OGLE-TR-111 \citep{tr-111}.
Targets were selected from the 73 OGLE transit candidates in
Carina using indications found in the light curves, such as transit
shape, radius of the eclipsing body, ellipsoidal variability, and/or
secondary eclipses, to reject 43 candidates that were almost certainly
binaries. The remaining 30 candidates were ranked as first or second
priority. All 14 first priority candidates were observed and 11 of the
16 second priority candidates were observed (OGLE-TR-70, 71, 73, 74,
and 115 were not observed). The RV data reveal nine
candidates with small stellar companions, one grazing eclipsing
binary, four triple systems, one quadruple system, and fourteen
systems that appear to be binaries but whose precise orbits were not
determined.  Two additional objects (OGLE-TR-124 and OGLE-TR-131)
showed no RV variations in phase with the transit period 
($P<2\,$days in both cases) 
despite 8 RV measurements each with typical errors of $\sim 60\,\ms$.
If they have VHJs, therefore, these must be less massive than
$0.3\,M_J$ at the $2\,\sigma$ level.
A third object, OGLE-TR-109, showed large rotational broadening
with correspondingly high RV uncertainties and no
variation within those uncertainties. However, it seems to have the
spectral temperature of an F-star, which makes the primary too large
to host a HJ or a VHJ as a secondary given the transit depth. Seven
more candidates possess no spectral lines strong enough for 
RV measurements. However, \citet{pont} show that all but one
object (OGLE-TR-82) cannot be isolated, slowly rotating late-type
stars and thus they probably do not host a planet. The spectrum of
OGLE-TR-82 had very little cross-correlation signal.

\subsection{Summary of Candidates
\label{sec:summarycand}}

A careful accounting shows
that, after imposing all of the selection criteria presented in 
\S\ref{sec:selcrit},
almost all of of the OGLE-III planetary transit candidates have been
either confirmed as planets or rejected.

Five of the candidates have proven to be
planets transiting their primary star with a period between 1 and
5 days. These systems are summarized in
Table~\ref{tab:planets}. The vast majority of the remaining candidates are
known to be binary star systems. Of those whose nature remains
unknown, two have no lines useful for RV followup
measurements (OGLE-TR-68 and 127), three have unknown periods
(OGLE-TR-43, 44, and 46), and 14 have periods longer than 5 days and
thus do not have a VHJ or a HJ in the period range relevant to the present
study. An additional 11 systems have transit depths greater than
0.04 magnitudes, while all detected planets have transit depths less
than 0.03 magnitudes and a typical value is more like 0.01 to 0.02
magnitudes. If any of these 11 candidates contain a planet, its
nature must be substantially different from the 5 detected.

At this point, 4 systems of unknown nature remain: OGLE-TR-51, 82, 
134, and 137. The properties of these systems are summarized in
Table~\ref{tab:unknown}. OGLE-TR-51 is quite faint, $I=16.7$.  From
OGLE-II photometry, we find $V-I=1.4$.  Hence, $V=18.1$, well above
our $V_{\rm max}=17.5$ limit.

\citet{pont} observed OGLE-TR-82, but the
spectrum was too noisy to determine its nature. The 2MASS catalog
indicates that this source is very red ($I-J=2.8$, $J-H=1.4$, based on
the OGLE $I$ band magnitude and the 2MASS $J$ and $H$ band
magnitudes). This red IR-based color is confirmed by a 
crudely calibrated
OGLE-III optical color $V-I=3.3\pm 0.2$.  Since $I=16.30$, this
implies $V=19.6\pm 0.2$, which is well beyond our limit and easily
explains the difficulties that \citet{pont} encountered obtaining a spectrum.

\citet{hh05} ``rule out'' OGLE-TR-137 because they derive a large
radius for the companion, $r= 0.167\pm 0.16\, r_J$ ($2\,\sigma$).
This is a reasonable argument if all undiscovered planets have radii
like those that have been found in the past.  However, given the still
exploratory nature of this field, one should adopt a more cautious
attitude.  

OGLE-TR-134 displays marginal evidence for ellipsoidal variations which,
if confirmed, would indicate that the transiting companion was a low
mass star rather than a planet.  However, no firm conclusion can be
drawn at present.

Until these two candidates are more thoroughly investigated, their status
remains open.  We handle the uncertainties about OGLE-TR-137 by restricting
our inferences about planets to those with radii $r<1.5\,r_J$.

If OGLE-TR-134 turns out to have a planet, the rates derived in this paper
will have to be revised upward.  However, this will be straightforward
given the tables that we provide.

\section{{Selection Effects}
\label{sec:selection}}

In our simulations, we will assume that genuine
transit light curves with $\alpha>11$ are selected.  However,
this assumption could in principle fail for either of two reasons:
the light curve failed the $SDE$ cut and therefore was never examined
by a human or it passed all cuts but was incorrectly rejected by
humans.  The false rejections can take one of two forms.
First, the algorithm could return the wrong period, leaving the
human to examine an incorrectly folded (hence garbled) light curve.
Second, the light curve might be too noisy for reliable by-eye
interpretation.  In total then, there are three potential problems, which
we examine in turn.

\subsection{{Correlated Errors: Effect of SDE Cut}
\label{sec:sde}}

\begin{figure}
\plotone{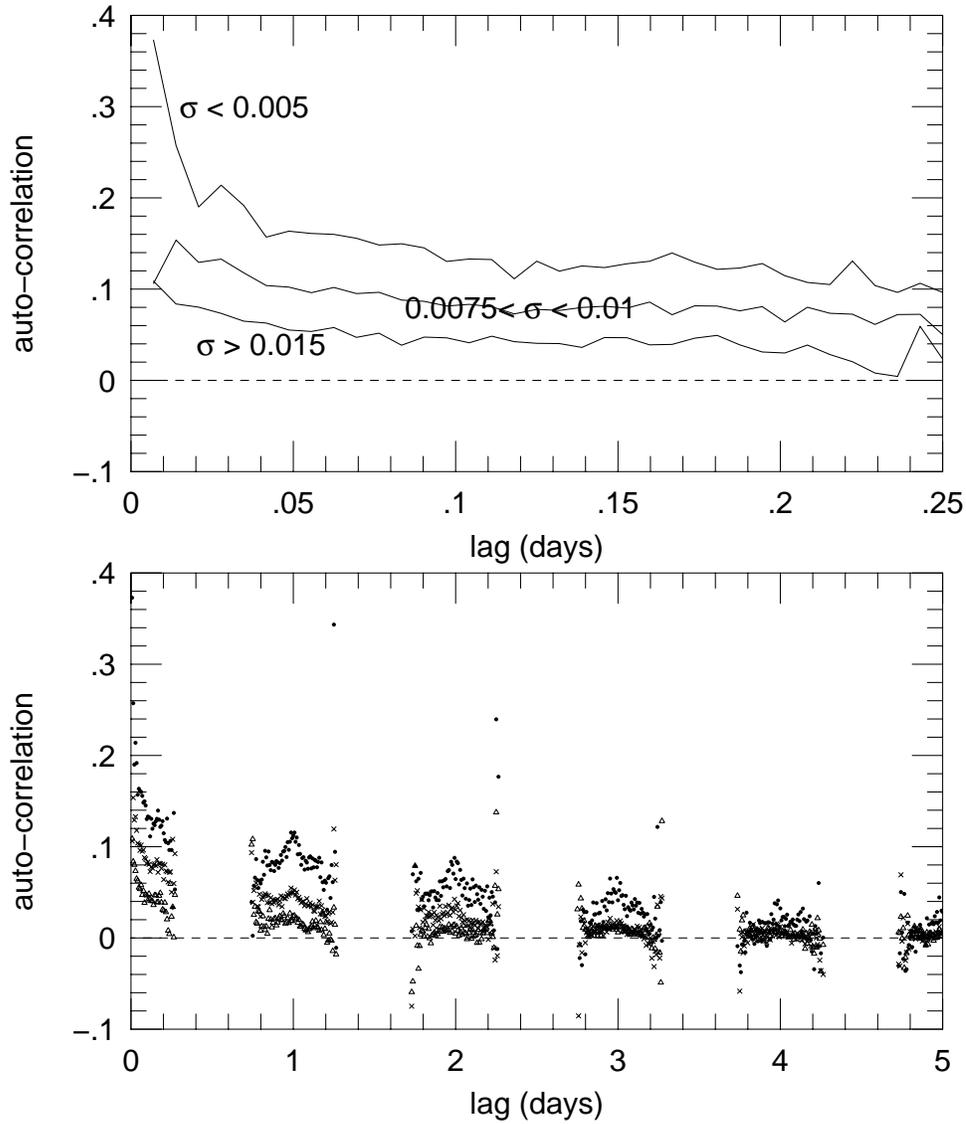}
\caption{\label{fig:autocor}
Autocorrelation function for transit data toward OGLE subfield
Car100.1.  The upper panel shows lag times ranging from 10 minutes to
6 hours for three subsamples of the data with photometric errors
(determined from the scatter) in the three indicated ranges.  The
lower panel shows data for these same subsamples over 5 days.
For lags similar to a transit duration (few hours) the autocorrelation
is significant, particularly at bright magnitudes (small errors),
which could in principle affect the detectability of the transits.}
\end{figure}

As discussed by \citet{pontsde}, OGLE errors are correlated, which tends
to drive down the $SDE$.  Figure \ref{fig:autocor} shows the autocorrelation
function for one of the 24 subfields in Carina.  That is, for each
lag time $\delta t$, we measure the correlation coefficient of all
pairs of measurements of the same star that are separated by $\delta t$.
For the stars with the smallest
errors ($\sigma<0.005$), the autocorrelation is extremely high (0.4)
at the shortest time lag of 10 minutes and drops to 0.16 at 1 hour.
While the autocorrelation is smaller for the more typical stars with
larger errors, it does not approach zero until lag times of 2--3 days.
Autocorrelation on the timescales of a transit (typically 
several hours) decreases the $SDE$.

\begin{figure}
\plotone{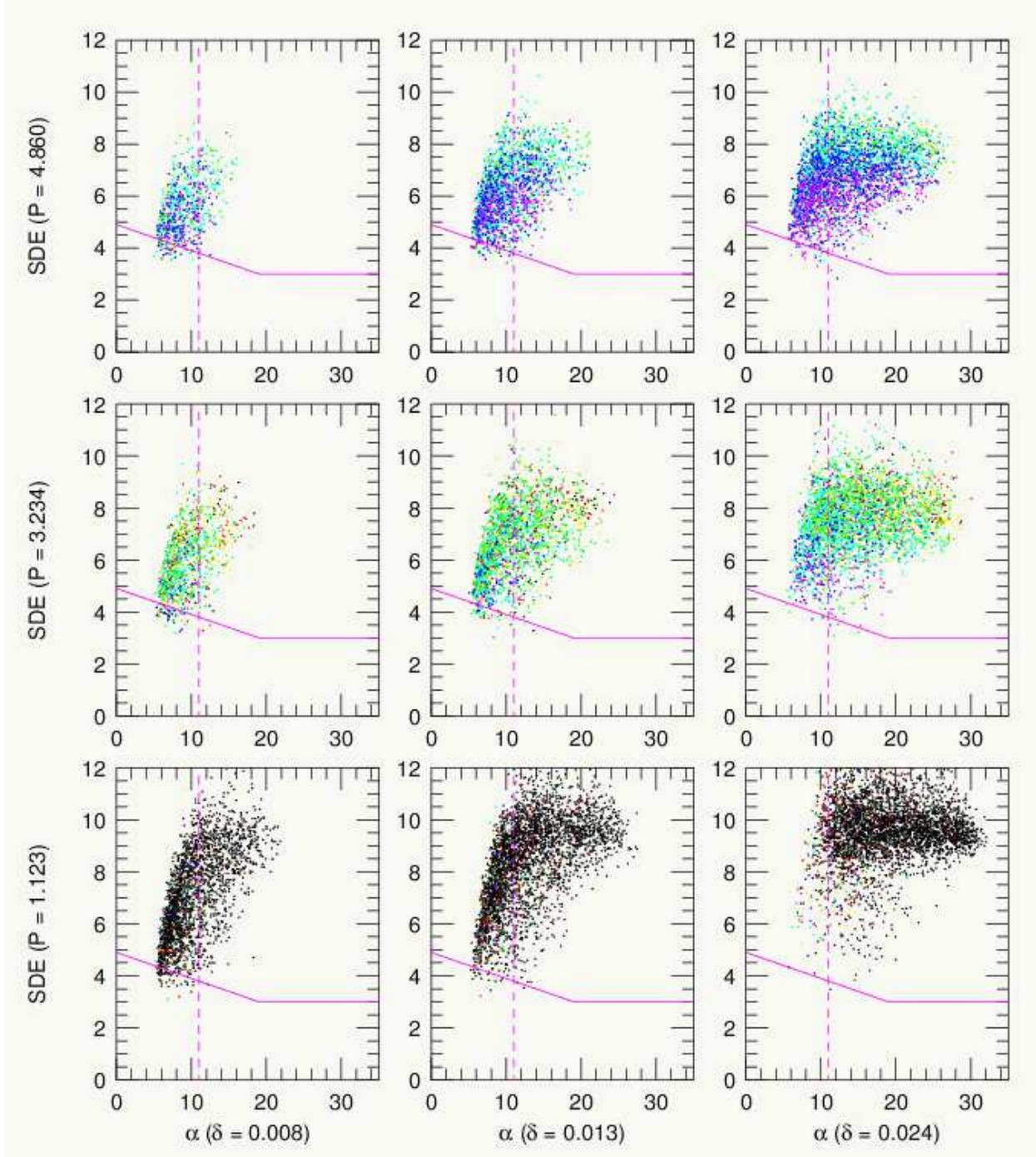}
\caption{\label{fig:alpsde}
Signal detection efficiency ($SDE$, measuring the strength of the
detection relative to the noise in the periodogram power spectrum)
vs.\ the signal-to-noise ratio parameter
$\alpha = [(S/N)^{-2} + N_{\rm obs}^{-1}]^{-1/2}$.
The nine panels show Monte Carlo simulations in which 
transits are injected into the data stream of each of the $\sim 4000$ stars
monitored in the OGLE-III Carina subfield Car100.1.  The
period $P$ and transit depth $\delta$ are held fixed at the indicated
values, while the phase and inclination of the transit are drawn randomly.
The selection criteria for $\alpha$ and $SDE$ are shown by magenta lines.
Essentially all simulated light curves that survive the $\alpha>11$ cut
also satisfy the $SDE$ cut, which means that it is not necessary to
model the $SDE$.
The points are color coded according to the number of distinct transits
that intersect the data stream: (magenta,blue,cyan,green,yellow,red)=(2--7),
while black = 8 or more.  At fixed $\alpha$, the $SDE$ tends to be lower for
light curves with fewer distinct transits.}
\end{figure}

To test the importance of this effect, 
we inject transits into the actual light curves
of stars in the same Carina field by subtracting flux from the
observations when a simulated planet passes in front of the star.
We then analyze the resulting light curve using the same \citep{kovacs} 
computer
algorithm that OGLE used to find transiting planets.  We consider
3 different periods ($P=1.123,3.234,4.86$ days), and three different
depths ($\delta = 0.008$, 0.013, 0.024).  The planet phase and orbital
inclination are drawn randomly.  The duration of the transit is
determined by assuming that the source is of solar mass and radius.
We remove all light curves for which the correct period is
not recovered to within 1\% because this problem is evaluated separately
in \S\ref{sec:period}. 
Figure~\ref{fig:alpsde} plots $SDE$ against $\alpha$.  The points are
color coded by the number of transits observed, ranging from magenta to
red for 2 to 7 transits, with black representing light curves with
8 or more transits.  As in the original OGLE-III analysis, 3 measurements
must be made during a transit for it to count as a separate transit. 
Note that there are extremely few points with
$\alpha>11$ and $SDE$ below the threshold, while there would be
many more such points were we to have adopted $\alpha>9$.  Note also
that the bluer colored points (fewer transits) tend to have lower $SDE$ 
at fixed $\alpha$,
a trend that is expected for correlated observational errors.
Finally, note that $\alpha$ saturates at $\alpha=\sqrt{N_{\rm obs}}\sim 33$,
as expected from equation (\ref{eqn:alphaeq}).

While very few transits are lost to the $SDE$ cut, Figure~\ref{fig:alpsde}
reveals another problem: there are a fair number of light curves with only
2 transits (magenta points), particularly at longer periods.  OGLE-III
eliminated these from its Carina sample, so we remove them from the model
as well.  
Based on the diagrams in Figure~\ref{fig:alpsde} as well as others not
shown, we find that the $SDE$ cut and 3-transit cut together eliminate
0.5\%, 1\%, 2\%, and 4\% of genuine transits for period intervals of 1--2 days,
2--3 days, 3--4 days, and 4--5 days, respectively.

\begin{figure}
\plotone{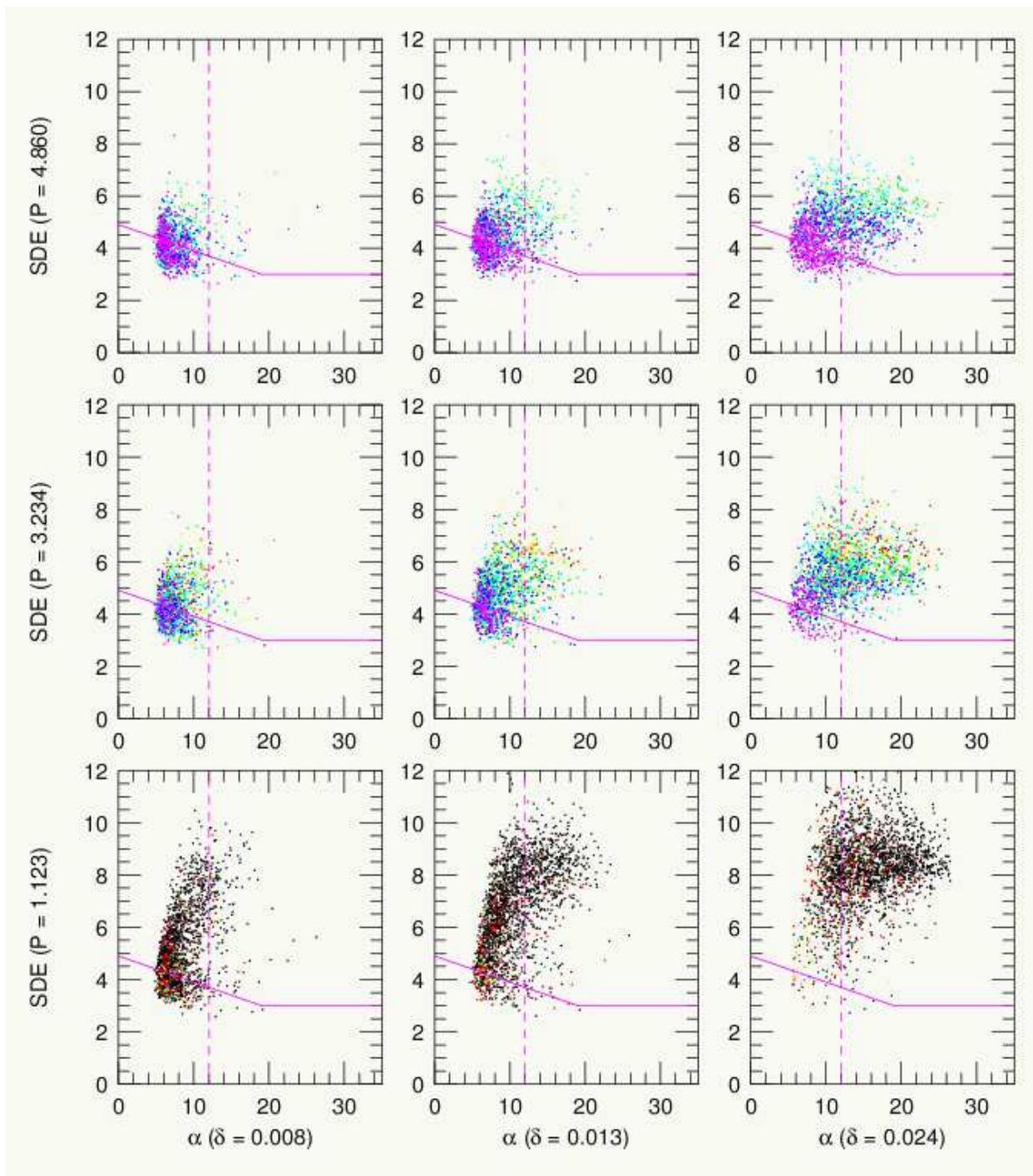}
\caption{\label{fig:alpsde2}
Similar to Fig.~\ref{fig:alpsde} but for OGLE-III bulge subfield BLG100.1.
For shorter periods $P\la 3\,$days, transit candidates that survive the
$\alpha$ cut also pass the $SDE$ cut (as was true toward Carina).  However,
at the longest periods, $P\sim 5\,$days, about 10\% of transits with
$\alpha>11$ fail the $SDE$ cut, a loss that must be taken into account
when determining planetary frequency.}
\end{figure}

Figure~\ref{fig:alpsde2} is a similar diagram for a bulge subfield.
In this case, there are some points that fall below the $SDE$ selection
threshold.  On the other hand, toward the bulge OGLE-III required
only 2 transits (as opposed to 3 toward Carina).  There are extremely
few single-transit light curves that survive the period-recovery requirement.
Toward the bulge we find that these two cuts eliminate 
0.5\%, 2\%, 4\%, and 9\% for the 4 respective period intervals.
We will take account of these casualties 
when we evaluate planet frequencies in \S\ref{sec:freqplan}.

\subsection{{Recovery of Transit Period}
\label{sec:period}}

Several thousand light curves (the great majority toward Carina)
passed the $\alpha$ and $SDE$ selection
criteria, and OGLE examined each of these by eye to determine whether
the transit was real.  The great majority of these were obviously
spurious, while a minority had to be studied more carefully to determine
whether they displayed real transits.  Since thousands of light curves
had enough periodic power to be selected without any transits, one should
ask whether the true period of any of the real transiting light curves was
overwhelmed by a spurious period.  If that occurred, then the
folded light curve would have had the same characteristics as typical
spurious transit light curves and most likely would have been rejected.

To address this question, we inject transits with five depths
\hfil\break\noindent
($\delta = 0.008,0.010,0.013,0.018,0.024$) into real light curves
as described in \S\ref{sec:sde}, but with random periods between
1 and 5 days.  We consider only light curves whose recovered parameters
meet our $\alpha$ and $SDE$ selection criteria.  We also demand that
the ``true'' value of $\alpha$ (based on the input parameters describing
the transiting planet) be at least 9.  We find that for virtually all
the simulated light curves, the recovered period either agrees to well within
1\% of the input period, or it disagrees by many percent.  We set the
threshold at 1\% and ask what fraction of light curves have ``bad periods''
by this standard.  We divide the sample into four period intervals,
1--2 days, 2--3 days, 3--4 days, and 4--5 days.  Toward Carina
we find that this fraction is roughly independent of depth 
and is about 6\%, 10\%, 13\%, and 20\%, in the four
period intervals.  Toward the bulge, the corresponding
percentages are 2\%, 9\%, 15\%, and 22\%.
We also take account of these losses
when we evaluate planet frequencies in \S\ref{sec:freqplan}.

\subsection{By-eye Selection
\label{sec:alphacomplete}}

One of us (AU) did the original by-eye selection of the OGLE-III
candidates and reported his subjective impression that this
process was complete for $\alpha>12$.  To objectively test this
impression, another member of our team constructed a sample of
61 light curves, 30 which contained artificial transits 
(injected into real light curves) with $11<\alpha<13$ and
periods of 3--5 days.  This is the most difficult period range
to vet because there tend to be fewer distinct transits, so the
correlated noise shown in Figure~\ref{fig:autocor} plays a bigger role.
The remaining 31 light curves were taken from the original data.
Both subsamples were selected from much larger ensembles of
respectively simulated and real light curves, using the same algorithm
to ensure that they displayed the broad range of distinct transits
shown in Figure~\ref{fig:alpsde}.  AU then applied the same
procedure to identify ``true'' transit candidates that he used to
construct the original sample.  He identified 33 light curves as
``real transits'' and the remaining 28 as ``spurious''.  The first
group contained all 30 simulated transits, plus 3 unmodified light
curves from the original sample.  All three of the latter had $\alpha$
near the bottom of the investigated range, $\alpha=11.1$, 11.2 and 11.4.
Comparison with the list of original candidates showed that two
of these had been identified in the original search (although one
was subsequently removed because it has only 2 transits), while the
third was not.  Of course, it is not known whether this final light curve,
which was missed in the original selection, is a real transit or not.
However, it is not qualitatively different in appearance from others
that are real.

The fact that 1 of 3 apparently real transits was missed in the
original search while all 30 of the simulated transits were identified in
the test indicates that the psychological conditions were
probably more difficult in the former.  This is plausible because
a relative handful of real transits had to be culled from many thousands
of spurious ones.

Based on these results, we estimate that the true effective threshold
(i.e., the location of the step function that is the best proxy for
the actual continuously declining completeness) is 
$\alpha_{\rm min} = 11.0 \pm 0.5$.  In \S\ref{sec:freqplan}, we will
evaluate the uncertainty in the final rates that is induced by this
uncertainty in the true value of the threshold.

\section{Modeling the Target Population
\label{sec:calculations}}

With the OGLE-III survey presumed
complete for $\alpha>11$, we are in a position
to examine the statistics of HJs and VHJs. We
would like to obtain the fraction of stars that have planets with any
given mass and orbital period. For this, we need to know the number of
systems that OGLE-III probed for planets of each period and mass. In
an ideal world we would determine, for each star in the survey, which
orbital periods and radii of planets would have been detectable had
such planets been present at transiting inclinations. We could then
assume that planets orbit at random inclinations relative to our line
of sight and obtain the number of systems probed by the survey for any
given kind of planet. The frequency of any type of planet is the
number of similar planets detected divided by the number of systems
probed for that kind of planet.

Unfortunately, this ideal approach is not possible because the
distances to most stars in the OGLE-III survey are unknown. We cannot
tell whether a given star is bright because it is close or because it
is luminous. This degeneracy makes it impossible to
determine whether a planet could have been detected around any
specific star in the survey, since whether or not a planet is
detectable depends on the star's luminosity, radius, and distance. To
circumvent this problem, we have created model star distributions for
the OGLE-III Carina and bulge fields based on the {\it
Hipparcos}~\citep{esa} stars within 50 parsecs of the Sun and a simple
Galactic model of stellar density. This model star distribution allows
us to statistically determine the number of systems probed, $N_p$, for
any kind of planet in the survey.

\subsection{{Star Distribution Model}
\label{sec:stardist}}

The Galactic model we use has a double-exponential disk with scale
height $h=250$ pc and scale length $H=2.7$ kpc, with the population at
the Galactocentric distance $R_0=8$ kpc fixed to fit observed local
values. The characteristics of the star population (number density as
a function of luminosity and radius) are derived from the {\it
Hipparcos} catalog. Each {\it Hipparcos} star is given a weight equal
to the inverse of the volume over which it could have been found.

{\it Hipparcos} is complete to visual magnitude
$V_{comp}(b)=7.3+1.1\left|\sin
b\right|$, where $b$ is Galactic latitude. For stars $M_V<3.805$, our
sample consists of all stars with distances $D<50$ pc and hence the
volume probed is ${\cal V}_{i}=(4\pi /3)(50 {\rm pc})^3$. For
$3.805<M_V<5$, we consider stars with $D<D_{7.3}$, where $D_{7.3}$ is
the distance at which the star would have a magnitude of 7.3, hence
${\cal V}_{i}=(4\pi /3)D_{7.3}^3$. The selection is more complicated
for the dimmest stars, $M_V>5$ for which we use
$D<D_{7.3+1.1\left|\sin b\right|}$, i.e., the limiting
latitude-dependent magnitude of {\it Hipparcos}, to define a
non-spherical volume in which the star could be found. 
$$
{\cal V}_i = \int_{-\pi/2}^{\pi/2} 2\pi 
d\sin b \int_0^{D_{7.3+1.1|\sin b|}} d D \,D^2
= {4\pi\over 3}\int_0^{\pi/2} d\sin b\,10^{0.6(12.3-M_V + 1.1\sin b)}
$$
\begin{equation}
= {4\pi\over 3}\,{e^\gamma-1\over \gamma}\,10^{0.6(12.3-M_V)}{\rm pc}^3,
\qquad \gamma \equiv 0.6*1.1*\ln 10.
\label{eqn:conehead}
\end{equation}
Each
star is assigned an absolute magnitude $M_V=V+5\log\pi-10$, where $\pi$
is the measured {\it Hipparcos} parallax in mas and $V$ is the {\it
Hipparcos} input magnitude. It is also assigned a radius $R_*=R_\odot
10^{0.597+0.536(B_T-V_T)-M_{V_T}/5}$ based on the prescription
of~\citet{gouldandmorgan}, where $B_T$ and $V_T$ are {\it Hipparcos}
pass bands and $M_{V_T}$ is the absolute magnitude in the $V_T$ band.

For any given star, $i$, the expected number of identical stars between
$D$ and $D+dD$ toward an OGLE-III field is given by
\begin{equation}
\frac{dN_i}{dD}=K\exp\left(-\frac{\left|z(D,b)\right|}{h}\right)
\exp\left(\frac{R_0-R(D,l)}{H}\right)\frac{\Omega D^2}{{\cal V}_i}
\end{equation}
\noindent where $(l,b)$ are the Galactic coordinates of the field,
$z(D,b)$ is the height above the plane at $(D,b)$, $R(D,l)$ is the
Galactocentric distance at $(D,l)$, $K$ is a normalization constant to
be described below, $\Omega$ is the solid angle covered by that field,
and $1/{\cal V}_i$ is the weight assigned to star $i$ that
characterizes its number density. The total number density of systems
probed for planets in a given field is found by summing $dN_i/dD$ over
all the stars in our {\it Hipparcos} sample.

An extinction model based on a double exponential Galactic dust disk
allows us to connect the three dimensional stellar density model to
the two dimensional fields observed on the sky.  We fix the dust scale
length at the scale length of the stellar Galactic disk, but we 
fit for the dust disk scale height, $h_{\rm dust}$. 
The local $I$-band extinction per unit distance,
$(dA_I/dD)_0$, sets the normalization of this disk. The ratio of
selective to total extinction, $R_{VI}$, is a fit parameter and is assumed
to be constant along the line of sight.
The following equation encapsulates our extinction model:
\begin{equation}
\label{eqn:extinction}
\frac{dA_I}{dD}=\left(\frac{dA_I}{dD}\right)_0 
\exp\left(-\frac{\left|z(D,b)\right|}{h_{\rm dust}}\right)
\exp\left(\frac{R_0-R(D,l)}{H}\right).
\end{equation}
\noindent The $I$-band extinction, $A_I$, to a given distance D is
then given simply by the integral of this equation. The $V$-band
extinction is $A_V=A_I R_{VI}/(R_{VI}-1)$.

To fix the model star density empirically in each OGLE-III
field, we produce a model binned color-magnitude diagram (CMD) and
vary four parameters to maximize the likelihood of obtaining the
observed OGLE-II CMD that lies along the OGLE-III field. We use
OGLE-II data because they have calibrated $V$ and $I$ photometry. To
produce the model CMD, we evaluate the color and magnitude at each
distance (including extinction) and add $dN_i(D,l,b)$ to the
appropriate color-magnitude bin for each star. The four parameters
varied are $h_{\rm dust}$, $(dA_I/dD)_0$, $R_{VI}$, and $K$, which is a
factor that normalizes the star density relative to the {\it Hipparcos}-based
value.

\begin{figure}
\plotone{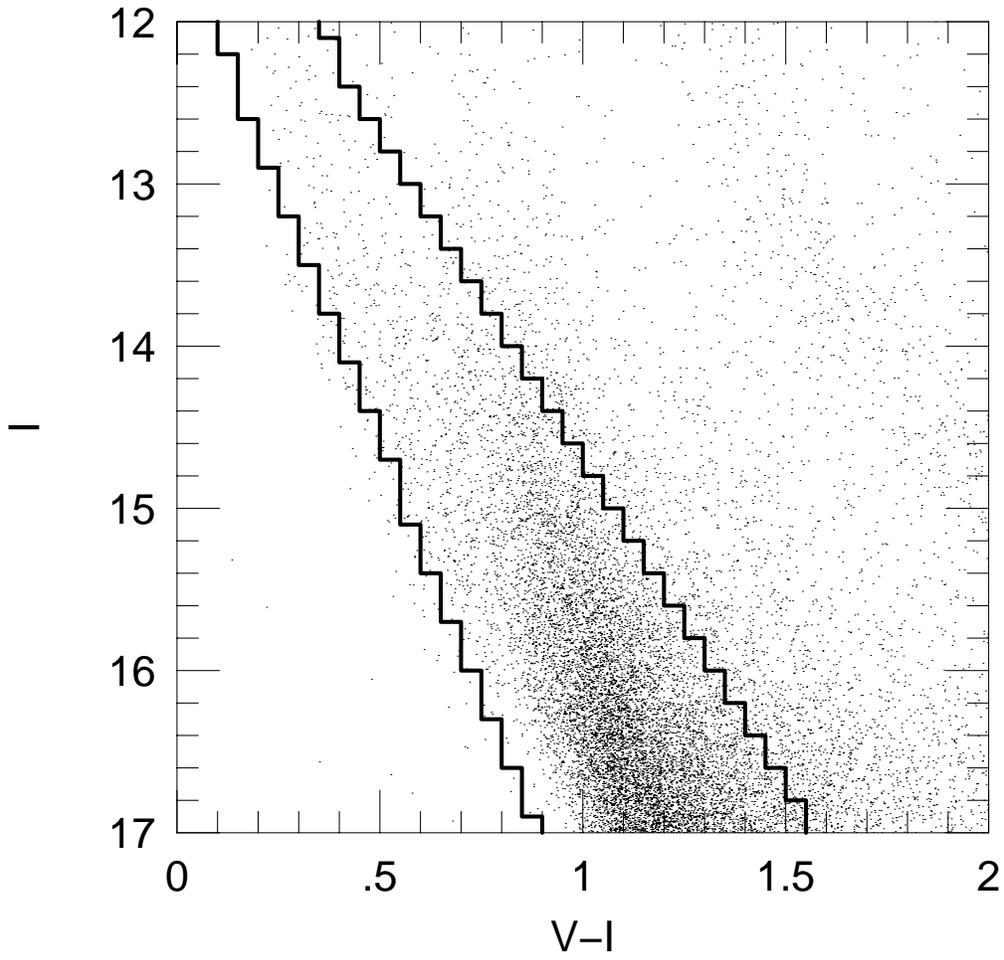}
\caption{\label{fig:3cuts} OGLE-II Carina CMD. The region inside the
bold histograms is used to fit a stellar model used along the OGLE-III Carina 
line of sight.  }\end{figure}

\begin{figure}
\plotone{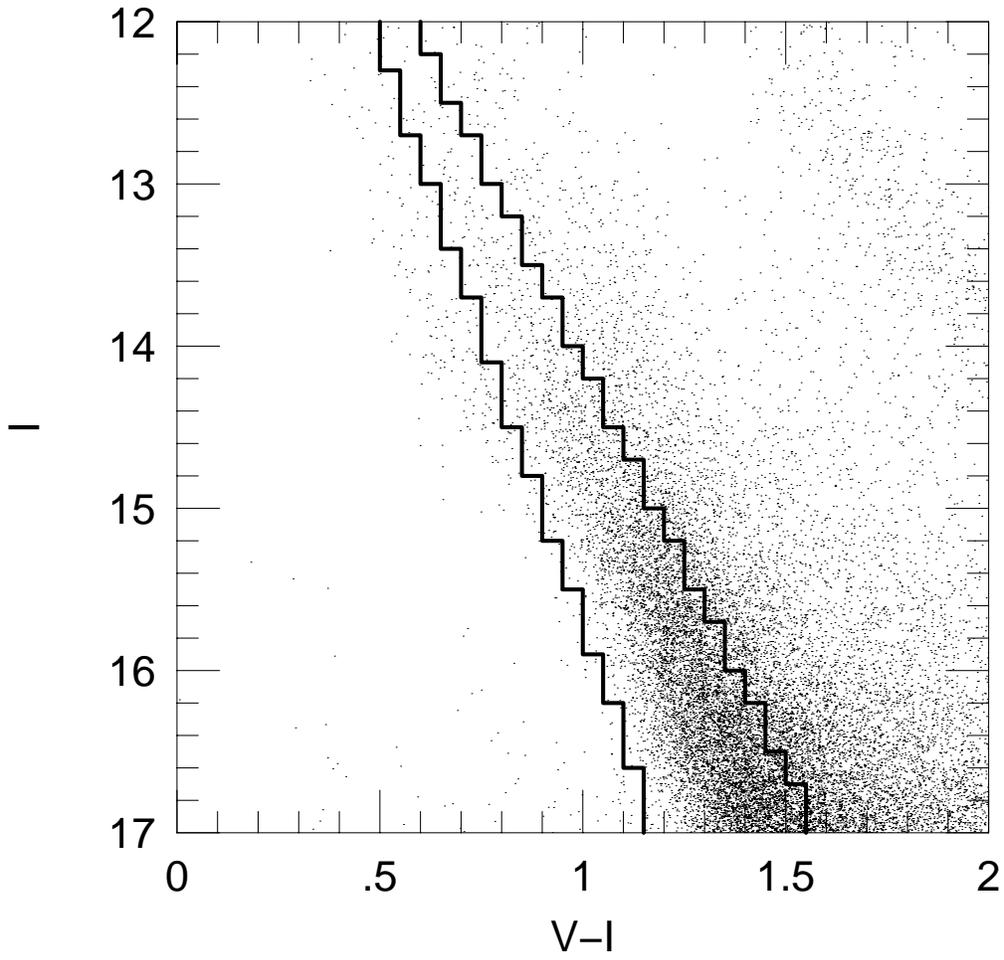}
\caption{\label{fig:bulge-cut} OGLE-II bulge CMD. The region inside
the bold histograms is used to fit a stellar model used along the OGLE-III
bulge line of sight.}
\end{figure}

We restrict the fit to the main sequence rather than the whole CMD
because we only expect planets to be detectable around main sequence
stars: giants have too large a radius, and the resulting 
magnitude change in transit would be too small to be detectable. To
implement this restriction, we make the cuts shown in
Figure~\ref{fig:3cuts} for the Carina field and in
Figure~\ref{fig:bulge-cut} for the bulge field.

\subsection{Resulting Fit Parameters
\label{sec:fitparms}}

\begin{figure}
\plotone{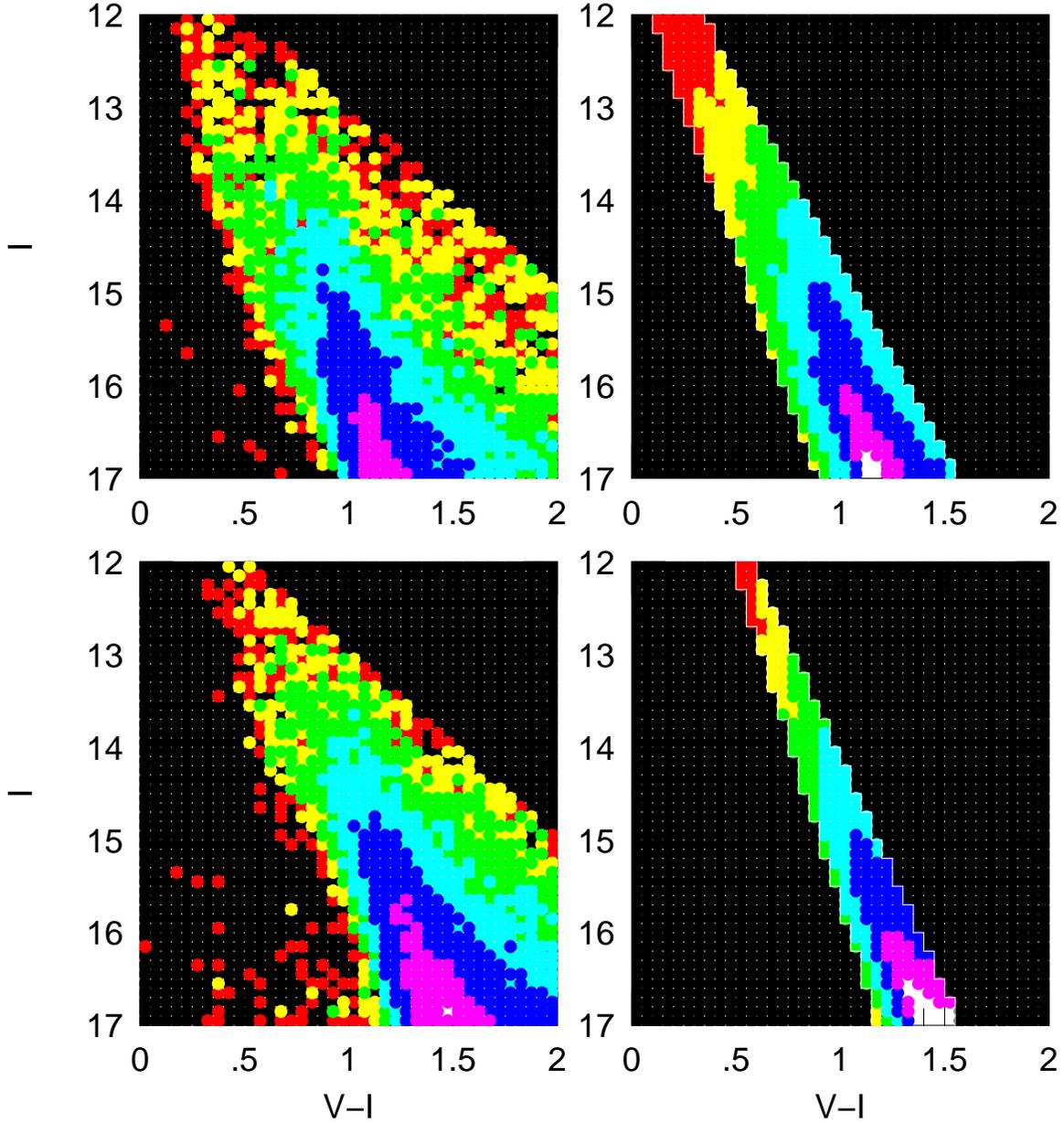}
\caption{\label{fig:carbul-obspred}
Observed (left) and predicted (right) CMD stellar densities $\rho$ 
for Carina (top) and the bulge (bottom).  
Contour levels in stars per ($0.05\times 0.1$) bin are
$\rho < 0.02$, 1, 3, 9, 27, 81, 140 for black, red, yellow, green,
cyan, blue, magenta,
and $\rho>140$ for white.  For Carina, the predictions are based on
the best fit model with parameters 
$(dA_I/dD)_0=0.174$ kpc$^{-1}$, $R_{VI}=2.23$, $h_{\rm dust}=273$ pc, and 
$K=0.81$, while for the bulge these parameters are
$(dA_I/dD)_0=0.435\,{\rm kpc}^{-1}$, $R_{VI}=2.56$, 
$h_{\rm dust}=65\,$pc, and $K=1.08$.
}\end{figure}

The top panels of Figure~\ref{fig:carbul-obspred} show the
OGLE-II color-magnitude diagram (CMD) for the Carina field and the
best-fit model CMD, each binned by both color and
magnitude. The whole observed OGLE Carina CMD is shown on the left, 
while the predicted CMD on the right shows only the region that
was used to fit the model to the OGLE
data. The lower panels are the
corresponding CMDs for the bulge fields. While the model CMDs
reproduce the grossest OGLE CMD characteristics in both fields, the
models and the observations are quite different in detail. This
difference indicates that our model is too simple. {\it Hipparcos}
stars may not be representative of OGLE stars. More likely, there are
some density characteristics along each of the Carina and bulge lines
of sight that are more complicated than a simple exponential disk. For
instance, the Carina fields may pass through a spiral arm, while we neglect
any such density fluctuations in our calculations.

To assess the impact of these discrepancies on our results, we evaluated
the number of stars probed for planets for a variety of other models
whose $\chi^2$ lies within 20 of the minimum (at the best fit).  We
find that these differ by of order 10\%, which is much less than
the Poisson uncertainties arising from the small number of planet
detections.

\section{Number of Systems Probed
\label{sec:numprobed}}

To establish the frequency of planets with a given radius and period,
we compare the observed planets to the calculated number of systems
probed for similar planets. To determine the latter, we
must combine the {\it Hipparcos}-based model of the stellar
distribution along the line of sight (\S\ref{sec:calculations})
with the selection criteria by which the detected transiting 
planets were actually culled (\S\ref{sec:planetids}).  The
most important of these is the signal-to-noise criterion, $\alpha>11$,
the modeling of which requires knowledge of the error distribution
of the OGLE-III data set.

\subsection{Error Functions
\label{sec:errorfunctions}}

We estimate the photometric errors of the OGLE-III light curves from their
scatter.  Of course, this estimate is accurate only if the star is
intrinsically stable: for variable stars the errors will be overestimated.
However, for the relatively few
non-transiting variables, the flux variations are a genuine
source of noise that would interfere with the detection of a transit
in a way that is very similar to photometric noise.  This is
because their more-or-less continuous variations have a very
low correlation with the set of periodic, short-duration flux decrements
that characterize the transit template used by the search algorithm.
On the other hand, the number
of transiting variables is so small that it has a negligible impact
on the assessment of the photometric errors of the target population
as a whole.  Hence, the photometric scatter in this population is indeed
the relevant quantity.

Our approach will be to estimate the photometric error as a function of
$I$-band magnitude for each of the two target directions, Carina and the
bulge.  We must also check that a single function reasonably describes
all the subfields of each of these directions.  We immediately confront
the problem that the OGLE-III data are not precisely calibrated.
Toward the bulge, each of the three fields overlaps a (calibrated) OGLE-II
field, which enables a calibration accurate to a few hundredths of a magnitude.
However, toward Carina, only one of the three fields overlaps OGLE-II.
For the other two Carina fields, we obtain approximate calibrations by
comparing OGLE-III flux levels to those of OGLE-III observations of OGLE-II 
fields made on the same photometric
night.  We expect these calibrations to be accurate to better than 0.1 mag.

\begin{figure}
\plotone{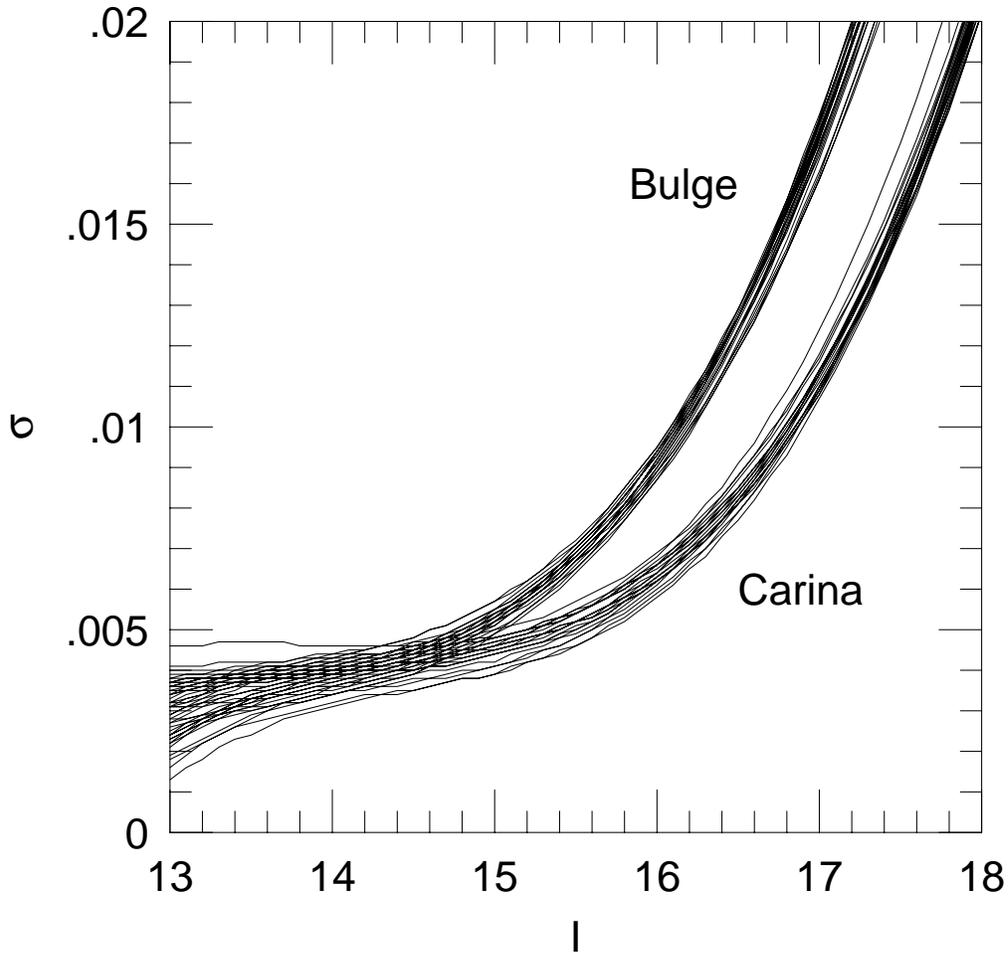}
\caption{\label{fig:errorfunctions}
Error functions vs. $I$ magnitude for each of 48 subfields, 8 for each
of three fields in each of two target directions (Carina and bulge).
Each is a cubic that is fit to the data after recursive elimination
of $2.5\,\sigma$ outliers.  The 24 error functions in each direction
are tightly clustered.  The bulge errors are worse at faint magnitudes
because of shorter exposures and higher background.}
\end{figure}

We separately analyze a total of 48 subfields, 8 subfields for each of
the three fields in each of the two target directions.  For each we
fit the errors, $\sigma$, to a cubic polynomial in the calibrated $I$ 
magnitude.
We eliminate the largest outlier and repeat the procedure until all 
$2.5\,\sigma$ outliers are eliminated.  The resulting 48 error functions
(Fig.~\ref{fig:errorfunctions}) are tightly clustered in two groups,
one for each target direction.  At the faint end, there is only one
significant outlier (a Carina subfield), and that by a modest amount.
While, there is some scatter in the error functions toward the bright end,
the ensemble in each target direction can be reasonably represented as a
single function.

\begin{figure}
\plotone{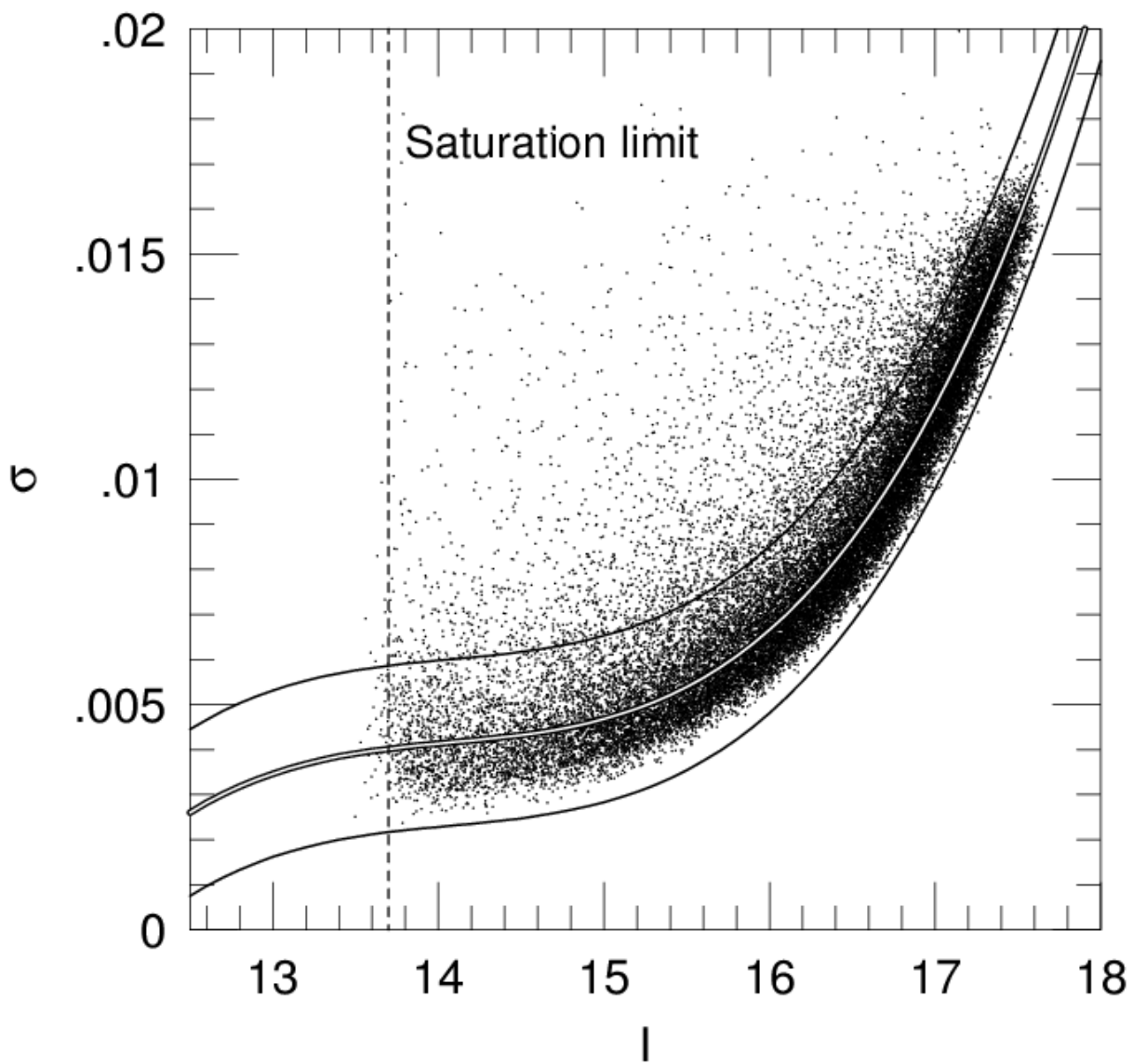}
\caption{\label{fig:carinaerrors}
Mean error function ({\it silhouette curve})
vs. $I$ magnitude for Carina.  Points are the errors
(determined from the scatter) of individual stars in CAR100, the
only one of the three Carina fields that can be directly calibrated.
The {\it solid curves} indicate the $2.5\,\sigma$ limits beyond which
data points were removed from the fit.  The 10\% of stars so eliminated
are assumed to contribute to planet detection with half the efficiency
of the 90\% that are well represented by the mean error function.
The dashed vertical line represents the saturation threshold, which
is identified from the points in this diagram to be at $I=13.7$.}
\end{figure}

\begin{figure}
\plotone{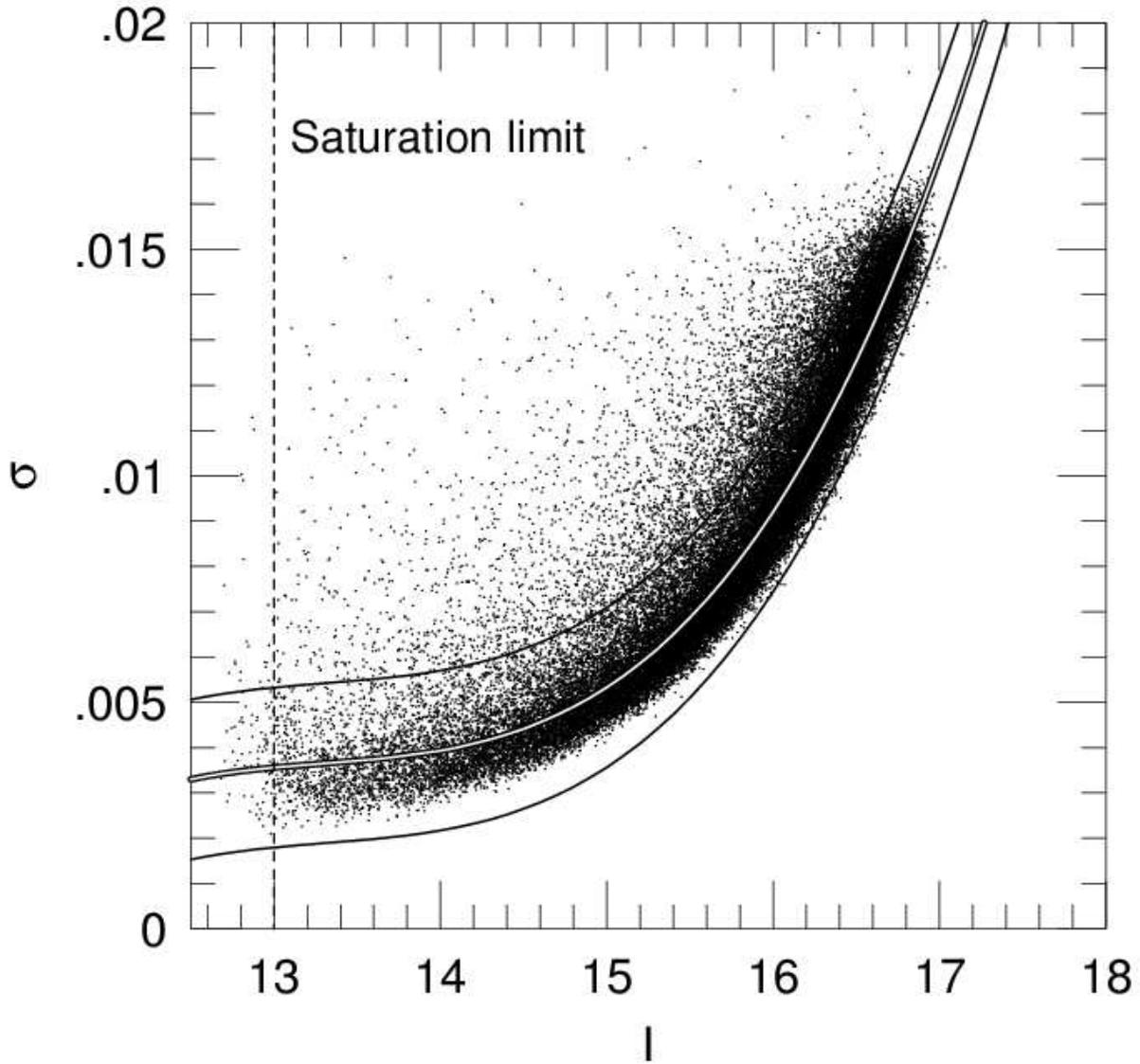}
\caption{\label{fig:bulgeerrors}
Mean error function ({\it silhouette curve})
vs. $I$ magnitude for the bulge.  Similar to Fig.~\ref{fig:carinaerrors}
except that in this case all three bulge fields are included in the fit.
Again, roughly 10\% of the stars are eliminated from the fit by the
$2.5\,\sigma$ criterion ({\it solid curves}).
The dashed vertical line represents the saturation threshold, which
is identified from the points in this diagram to be at $I=13.0$.}
\end{figure}

We therefore repeat the above error-fitting procedure for an ensemble in
each target direction.  Because all three bulge fields are directly 
calibrated from OGLE-II data, we combine all 24 bulge subfields for
this analysis.  However, for Carina, we use only the 8 subfields of
CAR100, the one Carina field that overlaps OGLE-II.  In 
Figures~\ref{fig:carinaerrors} and \ref{fig:bulgeerrors}, we show the
resulting error functions together with the photometric scatter of each
of the stars in the respective target directions.  At faint magnitudes,
the error functions closely track the peaks of the ridge lines in the
data points.  At bright magnitudes, they ride slightly above the ridge
lines, but this properly reflects the asymmetric distribution of the
data around the ridge line.  Each figure also shows the
$2.5\,\sigma$ limits, beyond which all stars were removed from the analysis.
In each case, about $10.3\%$ of stars have been eliminated, the 
overwhelming majority because their errors are too large.  To take
account of this asymmetry, we will consider that these 10\% of the
stars contribute to planet detection with half the efficiency of
the 90\% that are well represented by the error function.  This is
obviously a crude approximation, but the error that it induces is
more than an order of magnitude smaller than the Poisson errors, so a more
exacting approach is not warranted.

The coefficients of the two cubics ($\sigma = \sum_{k=0}^3 a_k I^k$) are
$$
(a_0,a_1,a_2,a_3) = (-0.72335462,0.15435485,-0.01094052,0.00025906)
\qquad {\rm Carina}
$$
\begin{equation}
(a_0,a_1,a_2,a_3) = (-0.57434586,0.12976481,-0.00973238,0.00024385)
\qquad {\rm bulge}
\label{eqn:cubiccoefs}
\end{equation}

We note that at faint magnitudes, the bulge fields have larger errors
than the Carina fields.  This is partly because they have shorter
exposures (120 vs.\ 180 seconds), and partly because the
dense star fields toward the bulge add to the background ``sky''.
The two target directions are more similar at bright magnitudes, where
the mean scatter falls to 0.005 and ultimately to 0.004 magnitudes in each
case.

\subsection{Maximum Distance for Planet Detection
\label{sec:maxdis}}

The $S/N$ of the transit in the light curve ultimately sets the maximum
distance at which a planet could be detected around a given star
($D_{\rm max}$). The depth of the transit and the photometric error
determine the $S/N$:
\begin{equation}
\label{eqn:signal}
S/N=\frac{\delta}{\sigma}
\sqrt{\frac{N_{\rm tr}}{N_{\rm obs}}\left(N_{\rm obs}-N_{\rm tr}\right)},
\end{equation}
where $\sigma$ is the average photometric error in magnitudes
associated with individual data points, $N_{\rm tr}$ is the number of data
points in the light curve during transits, $N_{\rm obs}$ is the total
number of data points in the light curve, $\delta=(2.5/\ln 10)(r/R_*)^2$
is the fractional depth of the transit in the light curve, $r$ is the
radius of the planet, and $R_*$ is the radius of the star. For a
randomly sampled light curve, $N_{\rm tr}=f N_{\rm obs}$, where $f$ is the 
fraction of the orbit spent in transit, 
\begin{equation}
f=\frac{1}{\pi}\arcsin\sqrt{\frac{R_*^2-b^2}{a^2-b^2}},
\end{equation}
$a$ is the semi-major axis of the planet's orbit, and
$b$ is the impact parameter of the planet on the star, $0<b<R_*$.
We address the potential complication of non-randomly sampled light curves
in \S\ref{sec:resonances}.

These equations implicitly assume that the
transits are box-like, i.e., they ignore the light-curve distortions
induced by limb darkening and ingress/egress.  In fact, 
we take both of these effects into account.
However, as we discuss in \S\ref{sec:limbdarkening}, the practical
impact of these effects is small while the expressions representing
them are cumbersome.  Hence, for purposes of exposition, we initially
ignore both effects.

As discussed in \S\ref{sec:planetids} and \S\ref{sec:alphacomplete}, 
we estimate that the survey
is complete to $\alpha_{\rm min}=11$.  To incorporate this constraint
into the simulation, we must relate the $(S/N)$ to $\alpha$, which
is defined analogously to equation (\ref{eqn:signal}) but with
$\sigma \rightarrow \sigma_{\rm scatter}$.  Because $\sigma_{\rm scatter}$
is determined from the light curve {\it without} fitting for the transit,
it is augmented relative to $\sigma$ by
\begin{equation}
\sigma_{\rm scatter}^2 = \sigma^2 + 
{\delta^2N_{\rm tr}(N_{\rm obs}-N_{\rm tr})\over N_{\rm obs}^2}.
\label{eqn:sigmascatter}
\end{equation}
Substituting this equation into equation (\ref{eqn:signal}) yields
equation (\ref{eqn:alphaeq}) for the relation between $\alpha$ and
$(S/N)$.  Since there are approximately 1120 and 800 observations in the
Carina and bulge fields, our adopted threshold of $\alpha_{\rm min}=11$
translates into
\begin{equation}
(S/N)_{\rm min}^{\rm Carina} = 11.647,
\qquad
(S/N)_{\rm min}^{\rm bulge} = 11.940
\label{eqn:snmin}
\end{equation}
For each
 {\it Hipparcos} star, 
we invert equation (\ref{eqn:signal}) using $S/N=(S/N)_{\rm min}$ to obtain
the maximum photometric error at which a planet could be detected
($\sigma_{\rm max}$). We then invert equation~(\ref{eqn:cubiccoefs})
to find the faintest $I$ mag at which the transiting planet could be detected.

Next, we dim each star for dust $I=I_0+A_I(l,b,D)$ according to equation
(\ref{eqn:extinction}). This extinction as a function of distance,
combined with the distance modulus and the absolute magnitude of the
star, allows us to convert from the maximum magnitude at which a
planet could be detected to the maximum distance at which that planet
could be detected ($D_{\rm max}$). We calculate the number of stars
($N_S$) around which that planet could be detected if it transited the
middle of the star ($b=0$):
\begin{equation}
\label{eqn:goodstars}
N_S=\sum_i \int^{D_{\rm max},i,b=0}_{D_{\rm min},i} dD \frac{dN_i}{dD}, 
\end{equation}
where $D_{{\rm min},i}$ is the minimum distance to which star $i$ can
be detected, which is set by the saturation limits shown in 
Figures~\ref{fig:carinaerrors} and \ref{fig:bulgeerrors}.
The number of systems probed for planets ($N_p$) by the
OGLE-III survey is significantly less than $N_S$ because planets orbit
stars at a random inclination to our line of sight, rather than edge
on.  Naively, this would imply an additional multiplicative 
factor of ${R_{*,i}}/{a}$,
where $R_{*,i}$ is the radius of star $i$ and $a$ is the semi-major axis
of the planet's orbit, to account for the probability of a transit. 
In fact, because $D_{{\rm max},i}$ is a function of transit duration (and so
of impact parameter), we must integrate over impact parameter 
from $b=0$ to $b=R_{*,i}$ for each star:
\begin{equation}
N_p=\sum_i \int_0^{R_{*,i}} {db\over a}\int_{D_{min,i}}^{D_{max,i},b} 
dD\frac{dN_i}{dD}.
\end{equation}
\citet{bulge,carina} remove a number of light curves from their sample
before running the transit search algorithm in order to eliminate
unlikely candidates with minimal effort. We include these cuts in our
calculations by modifying the maximum distance, instituting a minimum
distance, and discarding stars as appropriate. In both the Carina and
bulge fields, transit depths greater than 0.08 mag were discarded
prior to running the transit detection algorithm. In our calculations,
we discard transits with depths greater than 0.04 mag to match the set
of OGLE-III candidates we consider. Additionally, \citet{bulge,carina}
reject light curves with rms variations greater than 0.015 mag in all
fields.  We exclude such stars from our model, taking account of both
photometric errors and the additional scatter induced by the transit
itself.  In the bulge fields,
\citet{bulge} select light curves using a cut in the CMD 
in order to limit candidates to those on or near the main
sequence. Our calculations account for this cut by introducing a
minimum distance necessary to dim and redden the star to within the
accepted region. As mentioned above, the saturation threshold in each
field also imposes a minimum distance.
We then integrate from the minimum distance to the
maximum distance, rather than from zero to the maximum distance.

Photometric searches for planets by this technique are limited not
only by transit detection, but also by the feasibility of 
RV followup measurements. RV measurements are limited by
two factors: the $V$-band magnitude of the star (which influences the
$S/N$ of the spectrum) and the presence of spectral lines. In the
analysis that follows, we will restrict our attention to stars with
$(V-I)_0>0.4$ in order to eliminate early-F and A stars,
which have few spectral lines. The $V$-band magnitude limit for RV
measurements combines with the intrinsic color limit and an average
reddening to produce an apparent $I$-band magnitude limit. The
distribution of observed $I$-band magnitudes of the candidates that
proved susceptible to RV follow up suggests that RV
detections of transit candidates are complete to an effective
$I_{\rm max}=16.4$. From this, we infer a true $V$-band limit by adding
the $V-I$ limit and an average reddening (derived from our
simulations) of $A_V=0.7$ to obtain $V_{\rm max}=17.5$. In our
simulations, we implement these cuts by eliminating {\it Hipparcos} stars
that are too blue ($V-I<(V-I)_{\rm min}$) and by truncating our summation
for each {\it Hipparcos} 
star at the distance at which the star reaches $V_{\rm max}$.

\subsection{Number of ``Good'' Stars
\label{sec:goodstars}}

It has been widely noted \citep{gouldandmorgan,brown03,pont,gaudi05} 
that there seem to be far too few planets detected in the
OGLE-III survey based upon the planet frequency derived from RV
surveys. The argument for this apparent inconsistency goes as follows:
RV surveys found HJs orbiting roughly 1\% of all stars. At the
orbital radii under consideration, the geometric transit probability
is about 10\%. Since both the OGLE Carina and bulge fields contains
several tens of thousands of stars up to the limiting magnitude for
planet detection ($V_{\rm max}=17.5$, see $N_D$ in Table~\ref{tab:goodstars}),
one would naively expect about 40 planets to be detected in the Carina
field and about 80 planets to be detected in the bulge field. In
reality, 5 planets were detected between the two fields. 

We partially solve this seeming paradox based on data from our
simulations. We find that the OGLE-III transit survey is sensitive to
planets for only a small fraction of stars in the field. 
The number of stars probed
for planets at transiting inclinations, $N_S$, is given by
equation~(\ref{eqn:goodstars}), which indirectly depends on the limiting
visual magnitude, $V_{\rm max}$. Table~\ref{tab:goodstars} shows $N_S$ in
both the Carina and bulge fields for several different
$V_{\rm max}$. Table~\ref{tab:goodstars} also shows $N_D$, the approximate
total number of stars in the field that are brighter than
$V_{\rm max}$. Because the OGLE-III survey does not have secure $V$ magnitudes,
we scale the number of stars in a corresponding OGLE-II field by the
ratio of the solid angles occupied by the OGLE-III and OGLE-II fields
to obtain $N_D$. The ratio $N_S/N_D$ shown in
Table~\ref{tab:goodstars} gives the fraction of stars in the field
that are actually probed for transits. If the above reasoning is
applied to $N_S$ rather than $N_D$, one should expect to detect
9 planets versus the 5 actually found, which is not such a big
discrepancy.  Actually, a proper comparison (given next in
\S\ref{sec:freqplan}) must take account
of the precise period and planet-radii intervals probed.

At our adopted $V_{\rm max}=17.5$, only about 16\% of the stars
toward Carina and 4\% of the stars toward the bulge
can be probed for planets (of radius $r=1.2\, r_J$ and semimajor axis
$a=7.94\,R_\odot$).  These numbers are qualitatively similar to the
10\% found by \citet{gouldandmorgan} for transit surveys of bright
($V_{\rm max}=11$) stars.

Notice also that if the limiting followup magnitude were increased
from $V=17.5$ to $V=18$, the total number of systems probed would
only increase by 16\%.

Figure~\ref{fig:8} shows the number of systems probed based on our
stellar model for planets as a function of $M_V$, $(V-I)_0$ color,
distance from the Galactic center $R_{GC}$, and star radius
$R_*$ for the Carina and bulge fields.  These plots were produced
by applying the technique outlined in \S\ref{sec:stardist} (together
with the error function evaluated in \S\ref{sec:errorfunctions}) 
to the star-distribution
model with the best-fit parameters to the observed CMD.

In both fields, the distribution of stars probed peaks at early G stars,
as judged by either their color, absolute magnitude, or radius.
The stars typically lie about 1 kpc from the Sun: toward the bulge,
this places them at a Galactocentric distance of about 7 kpc, while
toward Carina, their Galactocentric distance is more similar to that
of the Sun.

\section{Frequency of Planets
\label{sec:freqplan}}

To make the most of sparse observational
data dominated by Poisson noise, we bin the 5 detected planets in
one-day-wide bins according to period. For example, for each period bin we
calculate the average number of systems probed for the range of periods 
within that bin but for the full range of radii (1--1.25$\,r_J$).
We convert the
period of the planet to the semi-major axis (used in our computations)
by assuming a stellar mass proportional to the radius for stars
smaller than the Sun and by assuming solar mass for stars with larger
radii. (To check the sensitivity of our results to this assumption, we
have also performed the calculations assuming all stars are solar
mass. The results agree to better than 10\%.) See
Table~\ref{tab:byperiod}.  In constructing this table, we take account
of the selection effects analyzed in \S\ref{sec:selection}.  
In particular, in each field we
reduce $N_p$ by the factors derived in \S\ref{sec:sde} to account
for transit light curves lost to the $SDE$ and minimum-transit-number cuts
and by the factors
derived in \S\ref{sec:period} to
account for the light curves whose periods are not accurately
recovered by the search algorithm.  Finally, we note that OGLE only
searched for periods $P>1.052$ days, so we restrict our simulation to
this range as well.

To establish the frequency of planets as a
function of radius, we perform a similar calculation, this time binned
by planet radius and integrated over the period range of 1--5 days.  
See Table~\ref{tab:byradius}.  In 
Figure~\ref{fig:logbyradius}, we show the sensitivity to planets
for each of the two fields, in each of two period ranges,
$1\,{\rm day} < P < 3\,{\rm days}$ 
and
$3\,{\rm days} < P < 5\,{\rm days}$.

From Table~\ref{tab:byperiod}, there were a total of 3 planets detected
in the 1--3 day period range from among 2062 stars probed,
and there are 2 planets detected from among 618 stars probed
in the 3--5 day range (for planet radii uniformly distributed over 
$1<r/r_J<1.25$).  These detections yield rates of 1/690 and 1/310
respectively.  Because of the small number of detections, the errors
in these estimates are completely dominated by Poisson statistics.
At 90\% confidence
we find that the frequency $F$ in these two ranges is
\begin{equation}
{1\over 1640} < F[1<P/{\rm day}<3] < {1\over 280} \qquad (90\%\, {\rm C.L.}),
\label{eqn:freqshort}
\end{equation}
\begin{equation}
{1\over 650} < F[3<P/{\rm day}<5] < {1\over 140} \qquad (90\%\, {\rm C.L.}),
\label{eqn:freqlong}
\end{equation}

From Table~\ref{tab:byradius}, the distribution of planet radii is
consistent with being flat between 1.00 and 1.25 $r_J$.  We have
sensitivity to planets with larger radii, up to $1.5\,r_J$,
but do not find
any.  For all radii $1.3<r/r_J<1.5$ we probe at least 1190 stars.
Hence, at 95\% confidence, we place an upper limit of $F<1/400$
for such large-radius planets in the period range of 1--5 days.
(We use 95\% one-sided limit here so that it can be directly
compared to the upper limit in the two-sided 90\% limits reported in
eqs.~[\ref{eqn:freqshort}] and [\ref{eqn:freqlong}].)

Our sensitivity to planets with $r<r_J$ is much weaker.  Nevertheless,
we can place a 95\% C.L.\ upper limit of $F<1/55$ for a model of planets 
uniformly distributed from 0.78 to 0.97 Jovian radii.

In \S\ref{sec:alphacomplete}, we estimated the completeness
threshold and uncertainty to be $\alpha_{\rm min}= 11.0\pm 0.5$.
This corresponds to a fractional error of 4.5\%.  By repeating
our analysis with $\alpha_{\rm min}=12$ rather than 11
(including estimating the reduced numbers of light curves lost
to the $SDE$ cut, bad period recovery, etc.) we find that the
survey sensitivity scales as $(S/N)^{-n}$ where $n=2.5$ for all
period intervals and $n$ ranges from 2.0 for $r=1.0\,r_J$ to 
3.0 for $r=1.25\,r_J$.  Hence, the 4.5\% uncertainty in the
true threshold for $\alpha$ translates into a 
$2.5\times 4.5\%\sim 11\%$ uncertainty in the sensitivity of the survey.
This is small compared to our statistical errors.

\subsection{Comparison with RV Detections
\label{sec:rvcomp}}

Are the rates reported here for HJs and VHJs consistent with the results
from RV surveys?  Two RV groups have presented data from which such
a comparison can be made. We begin by
examining the
catalog of RV planets assembled on the California \& Carnegie Planet Search
website 
http://exoplanets.org/almanacframe.html 
(last updated 6 Feb 2005) and restrict this
sample to planets that were detected in the survey of 1330 FGKM stars
described by \citet{marcy05}.  There are 15 stars harboring planets with
periods $P<5\,{\rm days}$: 
(HD73256, HD83443, HD46375, HD179949, HD187123b, HD120136, HD330075,
HD88133, BD-103166, HD75289, HD209458, HD76700, HD217014, HD9826b,
HD49674).  From the lack of references to the California \& Carnegie 
search team, three of these
(HD73256, HD330075, and HD75289) do not appear to be part of their
sample.  One star (HD88133) appears to be part of a separate sample
of metal-rich stars being surveyed by \citet{fischer05}.  This leaves
11 planets with periods in the range $2.985\leq(P/{\rm day})\leq 4.950$.
Even though one of these (HD83443) has a period that is slightly shorter than
3 days, it is fair to compare the entire group with the HJ transit sample.
However, not all of these planets could have necessarily been detected
by the OGLE-III transit survey.  The four lowest-mass planets
(HD49674, HD76700, HD46375, and HD83443) have masses 
$M\sin i/M_J = 0.11$, 0.19, 0.25, and 0.41, respectively.  In a sample of
11 planets with random inclinations, one expects one of these to have 
$\sin i \sim [1 - (10/11)^2]^{1/2} = 0.42$, so the appearance of HD83443 at
$M\sin i = 0.41\,M_J$ would not be unusual even if all the planets in this
sample were of roughly Jovian mass.  However, the fact that 3 of the planets
have $M\sin i \leq 0.25\,M_J$ indicates that they are probably genuinely
of substantially lower mass.  Thus, one may suspect that they are also
of lower radius, but this is far from certain.  
On the one hand, in the solar system all gas giants have roughly
the same density.  If that rule applied here, these planets would be
smaller by a factor 0.48 to 0.63.  Inspection of Table \ref{tab:byradius}
shows that the transit sensitivity drops off dramatically at these
radii.  On the other hand, both of the HJs found in the OGLE-III
survey (OGLE-TR-10 and OGLE-TR-111) have radii $r\sim r_J$ despite their
low mass, $M\sim 0.5\,M_J$.  See Table~\ref{tab:planets}.  If this
trend continued to still lower masses, then all 11 of the 
RV planets in the \citet{marcy05} survey would be accessible to the
OGLE-III transit survey.  To be conservative (in the sense of minimizing
the apparent conflict between the planet frequencies as determined by
the two methods), we adopt 8 as the number of the RV planets in the 
\citet{marcy05} survey of 1330 stars that could have been detected in the 
OGLE-III transit survey.  However, for completeness, we also report below
the results of a comparison based on that number being 11.
Any of the 5 planets detected
in the OGLE-III transit survey would have been detected by RV (had they
been orbiting stars in the \citealt{marcy05} sample) unless they were
at improbably low inclination angles.  Thus we should ask whether it
is statistically plausible that (planet-detections)/(systems-probed)
should be 2/618 for transits given that it is 8/1330 for RV,
assuming that the underlying frequency, $F$, is the same for both samples.

Since it appears naively that there are too few transit detections, we
formulate this question as: what is the probability that 2 or fewer
planets were detected given that $618 F$ were expected?  We average
over all $F$ weighted by its Poisson probability as determined from
RV and a uniform-log prior.  We find,
\begin{equation}
p_{\rm HJ} = {\int d\ln F\,P(n_1|N_1 F)\sum_{i=0}^{n_2} P(i|N_2 F)\over
\int d\ln F\,P(n_1|N_1 F)} 
=\biggl({N_1\over N_1+N_2}\biggr)^{n_1}
\sum_{i=0}^{n_2}
{(n_1+i-1)!\over (n_1-1)! i!}
\biggl({N_2\over N_1+N_2}\biggr)^{i}
= 34\%,
\label{eqn:hjrvst}
\end{equation}
where $n_1=8$, $N_1=1330$, $n_2=2$, $N_2=618$, and
$P(n,\tau)=\tau^n \exp(-\tau)/n!$.
This is certainly not low enough to demonstrate a conflict, but
one should nevertheless raise the question:  
Is there any reason to believe that the
transit and RV surveys sample different populations and, in particular,
that the transit sample should have a lower frequency of planets?

RV field samples are heavily biased toward metal-rich stars.  They
are selected on $B-V$ color and are essentially magnitude limited in $V$.  
At fixed $B-V$ color, $d M_V/d{\rm [Fe/H]}\sim -1.3$.  Hence, stars that
are more metal rich by $\Delta {\rm [Fe/H]}=0.3$ are over-represented in
the sample by a factor scaling as luminosity $L^{3/2}$ or
$10^{0.4*1.5*1.3\Delta {\rm [Fe/H]}} = 1.7$.  
As metal-rich stars are overabundant
in planets \citep{santos04,fischervalenti05}, the mean frequency of
planets in a magnitude-limited RV survey of field stars will be significantly
higher than the frequency in the underlying field population.

Transit surveys also suffer from this bias, but at a level that is about
an order of magnitude smaller.  Transit surveys are also magnitude-limited
and are also color selected, although the color for OGLE-III is $V-I$
rather than $B-V$.  However, (if we ignore extinction for the moment),
the effective depth of a transit survey scales as $L^{3/2}R^{-7/2}$
\citep{scaling} rather than $L^{3/2}$.  At fixed $V-I$ color (and
hence fixed surface brightness $S\propto L R^{-2}$), the depth therefore
scales as $L^{3/2}(L/S)^{-7/4}=L^{1/4}S^{7/4}$.  
Moreover, at fixed $V-I$, the scaling of $M_V$ with
metalicity is not as steep, $d M_V/d{\rm [Fe/H]}\sim -1.0$.  Hence,
for the above example of $\Delta {\rm [Fe/H]}=0.3$, the enhancement is
only $10^{0.4*0.25*1.0\Delta {\rm [Fe/H]}} = 1.07$.  Moreover, this
amplification is further diminished by the effects of extinction, which
undercuts the $L^{3/2}$ volume advantage of brighter sources.

Hence, while there is no statistical evidence for a lower frequency of 
transiting
HJs compared to RV HJs, we expect that such a trend may
emerge as the size of the sample increases.

If we assume that 11 (rather than 8) of the RV planets would have been
large enough to be detected in the OGLE-III survey, then 
equation~(\ref{eqn:hjrvst}) yields 17\% (rather than 34\%).  This
would be a more notable (though still not very significant)
discrepancy, but we again emphasize that it
is not clear that all 11 of the RV planets could have been detected.

We can also test whether the complete absence of RV planets with periods
$P<2\,{\rm days}$ is consistent with the three such planets detected
in the OGLE-III transit survey.  We again apply equation (\ref{eqn:hjrvst}),
but with $n_1=3$, $N_1=2062$, $n_2=0$, and $N_2=1330$, and find
\begin{equation}
p_{\rm VHJ} = \biggl({N_1\over N_1+N_2}\biggr)^{n_1} = 22\%.
\label{eqn:vhjrvst}
\end{equation}
Hence, the RV and transit surveys are marginally
consistent with respect to VHJs.

The Geneva group searched 330 stars for planets using RV, among which they
found 2 HJs (HD209458 and 51Peg) and no VHJs \citep{naef05}.  
Since the detections rates are very similar to those of 
California \& Carnegie Survey, but the sample is about 4 times smaller,
the results are even more consistent with the rates derived from
transits.

The Geneva group has also undertaken a much larger study, called ``CORALIE'',
which has surveyed 1600 stars and should have by this time detected
all HJs and VHJs in this sample.  However, to date they have reported
only 5 detections with $P<5\,\rm days$:
HD75289 ($P=3.51\,$day, $M\sin i=0.42\,M_J$), 
HD73256 ($P=2.55\,$day, $M\sin i=1.85\,M_J$), 
HD83443 ($P=2.99\,$day, $M\sin i=0.38\,M_J$), 
HD209458 ($P=3.52\,$day, $M\sin i=0.70\,M_J$) and
HD63454  ($P=2.82\,$day, $M\sin i=0.38\,M_J$).  See
\hfil\break\noindent
http://obswww.unige.ch/$\sim$naef/who\_discovered\_that\_planet.html.
If this is indeed the complete list of short-period
detections, the result is inconsistent at the 95\% confidence level
with the 11 such planets found by the California \& Carnegie Survey 
among 1330 stars (even assuming the two samples are disjoint).
Given this inconsistency, and given the fact that
the Geneva group has not yet commented on their apparent low rate of
detection, it seems premature to compare this result with the rates
presented here for transit detections.

\section{{Consistency Checks}
\label{sec:consistency}}

The results on planet frequency just derived depend critically on
the premise that the search for transiting planets was complete
over the parameter intervals assumed.  The three biggest worries
in this regard are:
\hbn{1)} The survey might not be truly complete all the way
down to our assumed threshold of $\alpha_{\rm min}=11$.
\hbn{2)} The sensitivity might be overestimated if the enhanced
correlation at bright magnitudes (low errors) seen in Figure~\ref{fig:autocor}
degraded OGLE's ability to recognize transits.
\hbn{3)} We may have misestimated the effective cutoff at faint
magnitudes imposed by the difficulty in obtaining spectroscopic followup
observations.

We discuss these concerns and our reasons for adopting the various
limits in question in \S\ref{sec:maxdis}. Nevertheless, it seems 
prudent to undertake additional tests.  To this end, we conduct two
Kolmogorov-Smirnov (KS) tests, one for the cumulative distribution
of the planets with respect to $S/N$ and the other with respect to $I$-band 
magnitude.  These are shown in Figures~\ref{fig:kssn} and \ref{fig:ksi},
respectively.  

If the survey were seriously incomplete at low $S/N$, we should expect a 
deficit of detections relative to the predictions shown at the right side of 
Figure~\ref{fig:kssn}.  No such deficit is seen.
If the errors were seriously underestimated at bright magnitudes, then we 
should see a  deficit at the left side Figure~\ref{fig:ksi}, while if 
the cutoff at faint mags had been misestimated, 
then there should be an excess or deficit of detections at the right side.
No such excesses or deficits are seen.  Formally, the KS
tests yield probabilities of $P(d)=0.50$ for $d=0.342$ and
$P(d)=0.82$ for $d=0.257$, respectively.

\section{Limb Darkening and Ingress/Egress
\label{sec:limbdarkening}}

Two effects cause transit profiles to deviate from the boxcar form
with which they are modeled in the \citet{kovacs} (hereafter BLS)
search algorithm:
limb darkening and ingress/egress.  Here we evaluate these 
two effects.  We give our prescription for including them in the
simulation, but also show analytically that their effects are 
relatively small.

We first consider limb darkening by itself.  In the linear approximation,
limb darkening is parameterized by a coefficient $\Gamma$ and
induces a surface-brightness profile (relative to uniform) of
\begin{equation}
S(x) = 1 - \Gamma\biggl(1 - {3\over 2}\sqrt{1-x^2}\biggr)
\qquad x\equiv {\theta\over \theta_*},
\label{eqn:sofx}
\end{equation}
where $\theta_*$ is the angular radius and $\theta$ is the angular
position on the surface of the star.  
Because BLS is fitting to a bimodal flux distribution, it will include
all points that have greater than half the mean depth as part of the transit
and will exclude all those with less than half the mean depth.  Hence,
if $\Gamma<(1+ 3\pi/8)^{-1}\sim 0.46$, it will find a transit length
equal to the chord traced by the center of the planet while it remains
inside the limb, i.e., $2(1-\beta^2)^{1/2}\theta_*$ where 
$\beta$ ($0\leq \beta \leq 1$) is the normalized impact parameter.
And it will find a ``depth'' $\delta$ equal to the
mean depth during this time interval, 
\begin{equation}
{\delta(\beta)\over\delta_0} ={\int_{-Q}^Q d y S(\sqrt{\beta^2 + y^2})\over 2 Q}
= 1 - \Gamma\biggl(1 - {3\pi\over 8}Q\biggr) 
\qquad Q\equiv \sqrt{1-\beta^2},
\label{eqn:lddepth}
\end{equation}
where $\delta_0$ is the depth for a uniform surface brightness.

For perfectly edge-on transits ($\beta=0$), the mean depth is fractionally 
higher by $0.18\Gamma$, while for grazing transits
($\beta\sim 1$), it is fractionally lower by $-\Gamma$.   Hence, edge-on
eclipses can be probed over greater volumes and grazing transits over
smaller volumes than would be the case for a uniform surface-brightness
source.  One may expect that these two effects would cancel out, and
we show below that this is approximately the case.  

However, before doing so we introduce ingress/egress effects.  These 
decrease the flux decrement (relative to the point-planet approximation)
when the center of the planet is near, but still inside, the limb of the
star, and increase it when the planet center is just outside the
limb.  However, from the standpoint of BLS, all information about the
latter effect is lost because the flux is closer to its non-transit level,
so the flux decrement is not included by BLS in its boxcar-approximation
``transit''.  To a first approximation then, BLS still finds a
transit length of $2Q$, but the mean decrement is reduced to
\begin{equation}
{\delta(\beta)\over\delta_0} = 
 1 - \Gamma\biggl(1 - {3\pi\over 8}Q\biggr) - (1-\Gamma){\epsilon\over Q^2},
\qquad \epsilon\equiv {2\over 3\pi}\,{r\over R_*},
\label{eqn:meandepth}
\end{equation}
where $r$ is the planet radius and $R_*$ is the star radius.  The various
factors in the final term can be understood as follows.  The ingress/egress
affects only the limb, where the flux is already reduced by a factor
$(1-\Gamma)$.  The length of the ingress and egress (restricted to the relevant
time when the planet center is inside the limb) is $2(r/R_*)/Q$,
hence its fractional length is $(r/R_*)/Q^2$.  Finally, the mean fraction
of the planet that is actually over the limb during these interval is
$2/3\pi$.  We use $\delta(\beta)$ as given by equation (\ref{eqn:meandepth}) 
in our simulations rather than the uniform surface-brightness value, $\delta_0$.
Since the stars probed by the OGLE-III survey are primarily solar-type
stars, we use the solar limb-darkening value $\Gamma_I=0.368$.  (This
corresponds to $c=3\Gamma/(\Gamma +2)=0.423$ in the usual limb-darkening
formalism, in which $S(x)\propto (1- c[1-(1-x^2)^{1/2}])$.)

In fact, during the time when the planet center is extremely near
(but still inside) the limb, the flux will be closer to its non-transit
than its mean-transit value, and BLS will assign such points to the
non-transit part of the light curve, thus shortening the fit duration
below $2Q$ and increasing the mean decrement relative to equation 
(\ref{eqn:meandepth}).  This will lead to very slightly lower $\chi^2$ 
and so to very
slightly higher $\alpha$ than implied by equation (\ref{eqn:meandepth}).
However, the effect is second order in $\epsilon$.  As we show immediately
below, the first order $\epsilon$ effects are already quite small,
so we ignore these second order effects in the interest of simplicity.

We find numerically that the net effect of including both limb darkening
and ingress/egress reduces the sensitivity of the OGLE-III survey by
5\%--2\%, for periods from 1 to 5 days.  However, it is also instructive
to understand this result analytically.  

Suppose that near the threshold of a survey, the number of systems
probed scales as $(S/N)^n$.  For example, in the uniform-stellar-density,
above-sky photon-limited noise case considered by \citet{scaling},
$n=3$, while below sky this would be $n=3/2$.  Then the total number
of systems probed (relative to ignoring limb darkening and ingress/egress)
would be
\begin{equation}
\eta(\Gamma,\epsilon;n) = 
{\int_0^1 d\beta [\sqrt{Q}\delta(\beta)/\delta_0]^n \Theta[\delta(\beta)]
\over \int_0^1 d\beta Q^{n/2}},
\label{eqn:Fn}
\end{equation}
where $\Theta$ is a step function.
In Figure~\ref{fig:ld}, we evaluate this numerically for a broad range 
$n$ and for $r/R_*=0$, 0.075, 0.1, 0.125, and 0.15.  

By comparing results of simulations with $\alpha_{\rm min}=11$ and 
12, we determine for the OGLE-III data set, that overall $n\sim 2.5$, and
ranges from 2 to 3 depending on what subclass of periods or radii are
examined.  Hence, the intersections of the curves with the dashed
vertical line in Figure~\ref{fig:ld} are the most
representative of the role of limb darkening and ingress/egress in the
present case.

Over the range of power laws shown in Figure~\ref{fig:ld}, 
limb darkening by itself affects the rate
at the level of a few percent, and is generally positive and gently
rising in the neighborhood of $n=2.5$ (relevant for the OGLE-III survey).
On the other hand, ingress/egress affects the rate negatively, also
by a few percent.  The net effect at $n=2.5$ is negative at the 2--4\%
level, in agreement with what we found by directly inserting limb darkening
and ingress/egress into the simulations.

\section{Periodic Observations
\label{sec:resonances}}

In the simulations reported in \S\ref{sec:calculations}, we
assumed that the observations were taken at random times and hence
that the fraction of all observations that occur in transit is exactly
equal to the fraction of the orbit that the planet is transiting the
star. However, real ground-based observations are not randomly
sampled, but rather can only be made at night. Depending on the phase
of the transit relative to the night, planets with periods close to an
integer number of days will be seen in transit with a frequency that may be
either much larger or much smaller than the average value we have
used. If the number of observations in transit is greater than
average, then the transit $S/N$ for a given stellar type and given
distance is also greater than average. Hence, the transiting planet
can be detected to a greater limiting distance. Conversely, if the
number of observations in transit is below average, then transits of
stars of a given type will be detectable to a smaller distance.

For observations at a fixed ratio of transit depth to photometric
error (but averaged over planets in all orbital phases at the
start of the observations), these divergent behaviors can lead to
a net enhancement or a net degradation of the survey sensitivity
to planets near integer-day periods \citep{pont,gaudi05}.
However, for searches extending to a given S/N limit, the effect can
only be an enhancement (assuming a uniform density of sources and
that extinction is not extremely severe).  Before presenting our
quantitative results, we examine this qualitative difference.

At sufficiently small transit depths (relative to photometric error)
and for non-special periods,
there will normally be too few observations in transit to yield
enough S/N for a detection.  However, near integer-day periods
(and for some phases), the number of transit observations
will increase by enough to meet the S/N threshold.  Hence, the
sensitivity will show upward spikes at integer periods.  Conversely,
at large transit depth, there will normally be more than enough
observations to meet the S/N threshold, but near integer-day periods
and for adverse phase alignment, there will be a shortfall.  Hence,
there will be negative spikes.  Since transits are easier to
detect at short periods, these two effects imply that there will
be negative spikes at short periods and positive spikes at long
period.  All these effects are shown in Figure 12 of \citet{pont}
and Figure 4 of \cite{gaudi05}.

However, these effects combine in quite a different way when considering
a survey to fixed limiting S/N.  Whether one is close to or far from
an integer period, the {\it mean number} of transits (averaged over
all phases at the start of observations) is the same.  Hence, near
an integer-day period, some phases have an amplified number of in-transit 
observations at the expense of
other phases.  Since (in the absence of extinction and with 
photon-limited above-sky errors), the maximum distance scales very
nearly as the square root of the number of in-transit observations,
the volume gained by the favorable phases exceeds the volume lost by
the unfavorable ones.  Hence, there is always a net gain in sensitivity
near integer-day periods.

To be more quantitative, we define the amplification factor 
$F_{amp}=\bar{N_p}(P)/N_{p,random}(P)$ as
the ratio of the phase-averaged number of systems probed in a model
ground-based observation scenario, $\bar{N_p}(P)$, of a planet with a
period $P$ to the number of systems probed in a randomly sampled
observation scenario, $N_{p,random}(P)$. We derive $F_{amp}$ from scaling
relations. Assuming a constant density of stars in space (neglecting
Galactic structure), $N_p$ is proportional to the volume of space within the
maximum distance to which a transit could be detected, ${\cal V}=d^3$
($N_p\propto d^3$). Neglecting extinction, the flux is proportional to $d^{-2}$,
while the photometric error of an individual data point $\sigma$ is proportional
to $({\rm flux})^{-1/2}$. Combining these scaling relations with the 
definition of
$F_{amp}$ and equation~(\ref{eqn:signal}) gives,
\begin{equation}
\label{eqn:A}
F_{amp}=\frac{1}{Pf(1-f)}\int_0^P \left[\frac{N_{\rm tr}(\phi,P,f)}{N_{\rm obs}}
\left(1-\frac{N_{\rm tr}(\phi,P,f)}{N_{\rm obs}}\right)\right]^{3/2} d\phi,
\end{equation}
\noindent where $\phi$ is the phase, $f$ is the fraction of the orbit
the planet spends in transit, $N_{\rm tr}(\phi,P,f)$ is the number of data
points in transit, and $N_{\rm obs}$ is the total number of data
points. We calculate $N_{\rm tr}(\phi,P,f)$ using a program that assumes
24-hour days, six hour nights, ten minute observations, an 80-day
observing campaign, and $f=(1/30)
(P/4\,{\rm day})^{-2/3}$. Figure~\ref{fig:A} shows $F_{amp}$ as a function
of period. 

Using the data shown in Figure~\ref{fig:A}, we find that
the average $F_{amp}$ for
periods from 1.5 to 4.5 days is 1.027. Thus, the net effect of periods
resonant with the observing cycle on the
calculated number of systems probed is $<3\%$. 
To assess the impact of
resonances on the distribution of periods that are actually observed,
we calculate the fraction of the total number of systems probed found
in the integer peaks (second row of Table~\ref{tab:peakfrac}) as
compared to the fraction expected without amplification based solely
on the widths of the peaks (first row of Table~\ref{tab:peakfrac}). 
For purposes of this calculation, we consider a ``peak'' to 
extend over the region within the interval $(n\pm 0.05)\,$days
where the local enhancement is at least 5\%.  We
find that integer-day resonances are not substantially more likely to
be detected than other periods in the period range explored here, 
which is consistent with
the planets actually detected. Since the net effect of resonances is
small and resonant periods are not much more likely to be observed
than non-resonant periods, we neglect integer-day resonances in our
calculations.

\section{Unresolved Binaries
\label{sec:unresolved}}

\begin{figure}
\plotone{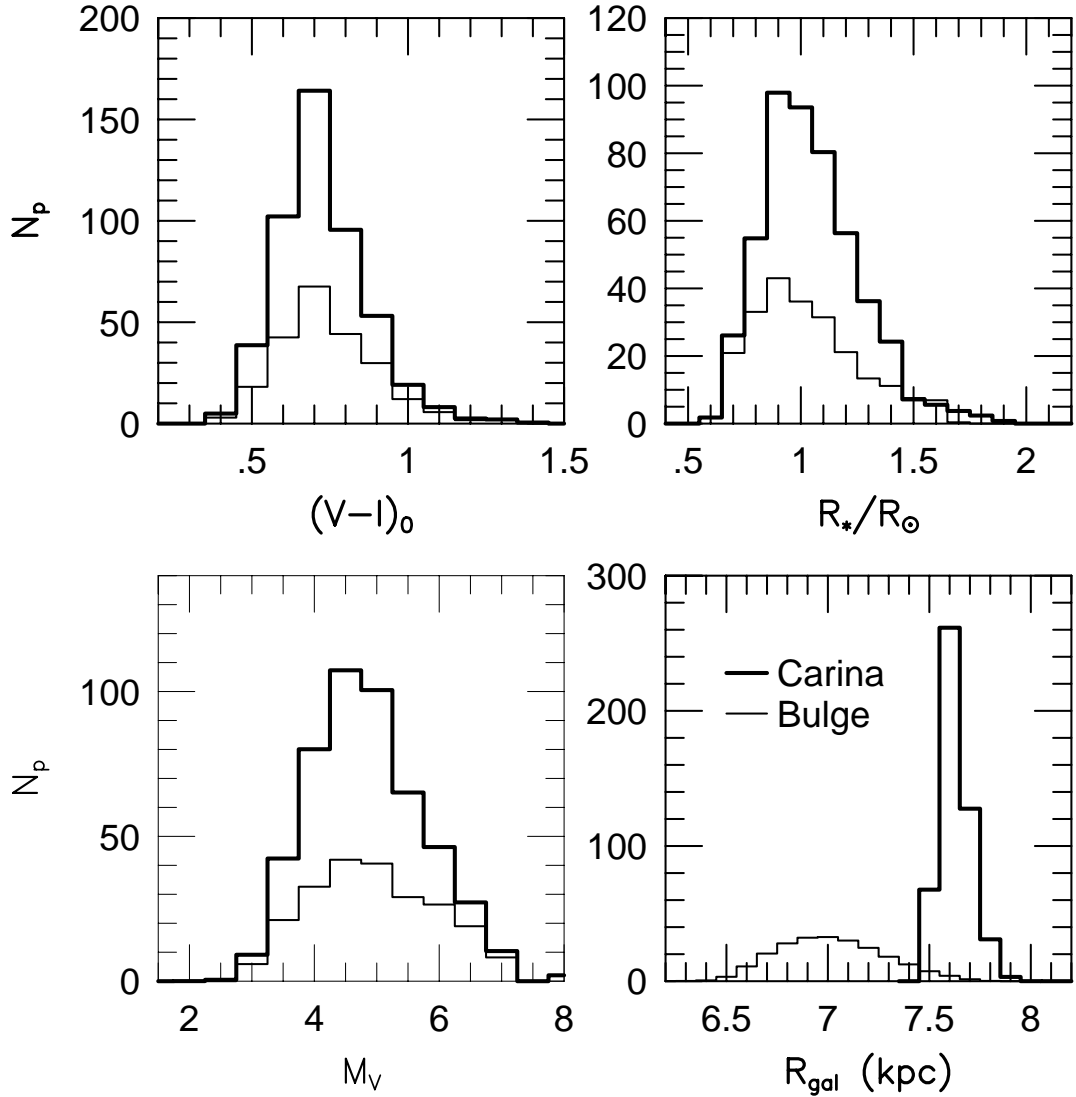}
\caption{\label{fig:8} Distributions of four properties of the stellar
populations probed for planets of radius $r=1.2\,r_J$ and $a=7.94\,R_\odot$
toward the Carina ({\it bold histograms}) and bulge ({\it solid histograms})
fields. Shown are the dereddened $(V-I)_0$ colors, the stellar radii $R_*$,
the absolute magnitudes $M_V$, and the Galactocentric distance $R_{\rm gal}$.
The Galactic-model parameters are $(dA_I/d I)_0=0.174\,{\rm kpc}^{-1}$, 
$R_{VI}=2.23$, $h_{\rm dust}=273\,{\rm pc}$,
and $K=0.81$ for Carina and
$(dA_I/d I)_0=0.435\, {\rm kpc}^{-1}$, $R_{VI}=2.56$, 
$h_{\rm dust}=65\,{\rm pc}$,
and $K=1.08$ for the bulge.}
\end{figure}

\begin{figure}
\plotone{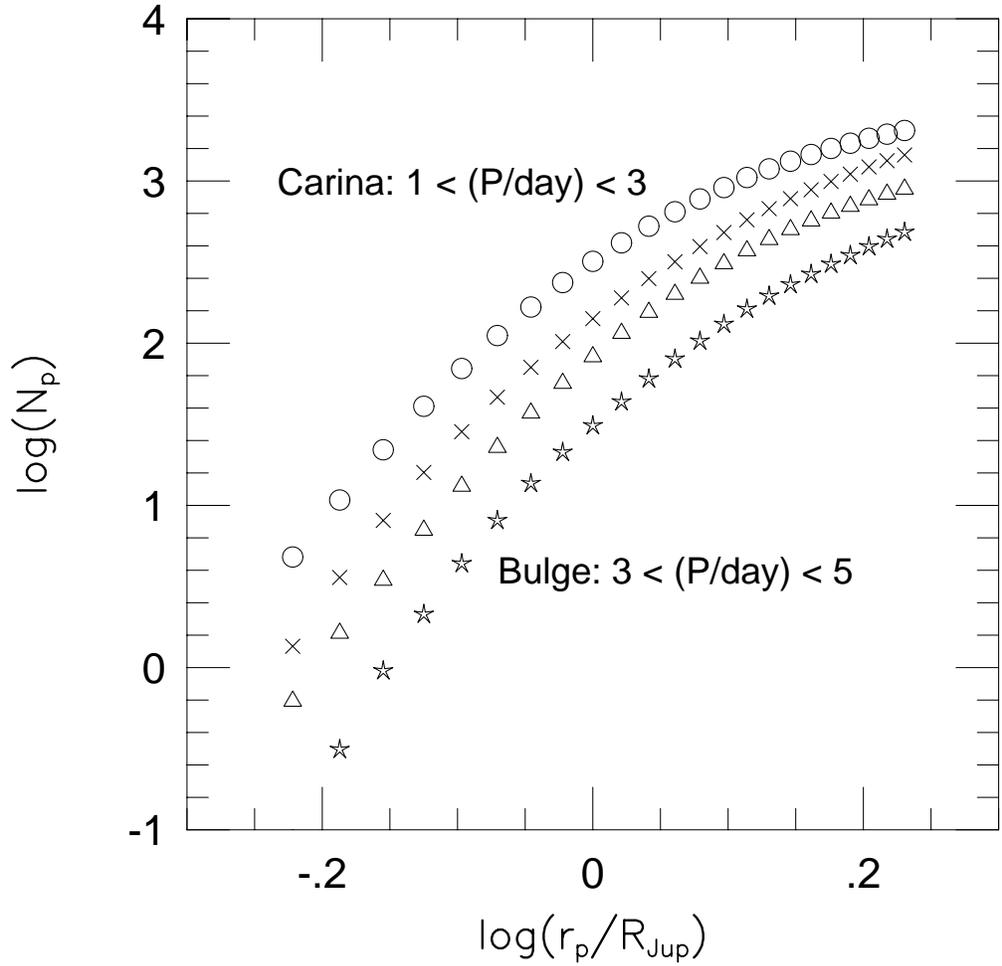}
\caption{\label{fig:logbyradius}
Number of systems probed $(N_p)$ as a function of radius averaged over a range
of periods. 
Carina: $1<(P/{\rm day})<3$ ({\it circles}),
$3<(P/{\rm day})<5$ ({\it triangles}).
Bulge: $1<(P/{\rm day})<3$ ({\it crosses}),
$3<(P/{\rm day})<5$ ({\it stars}).
}
\end{figure}

\begin{figure}
\plotone{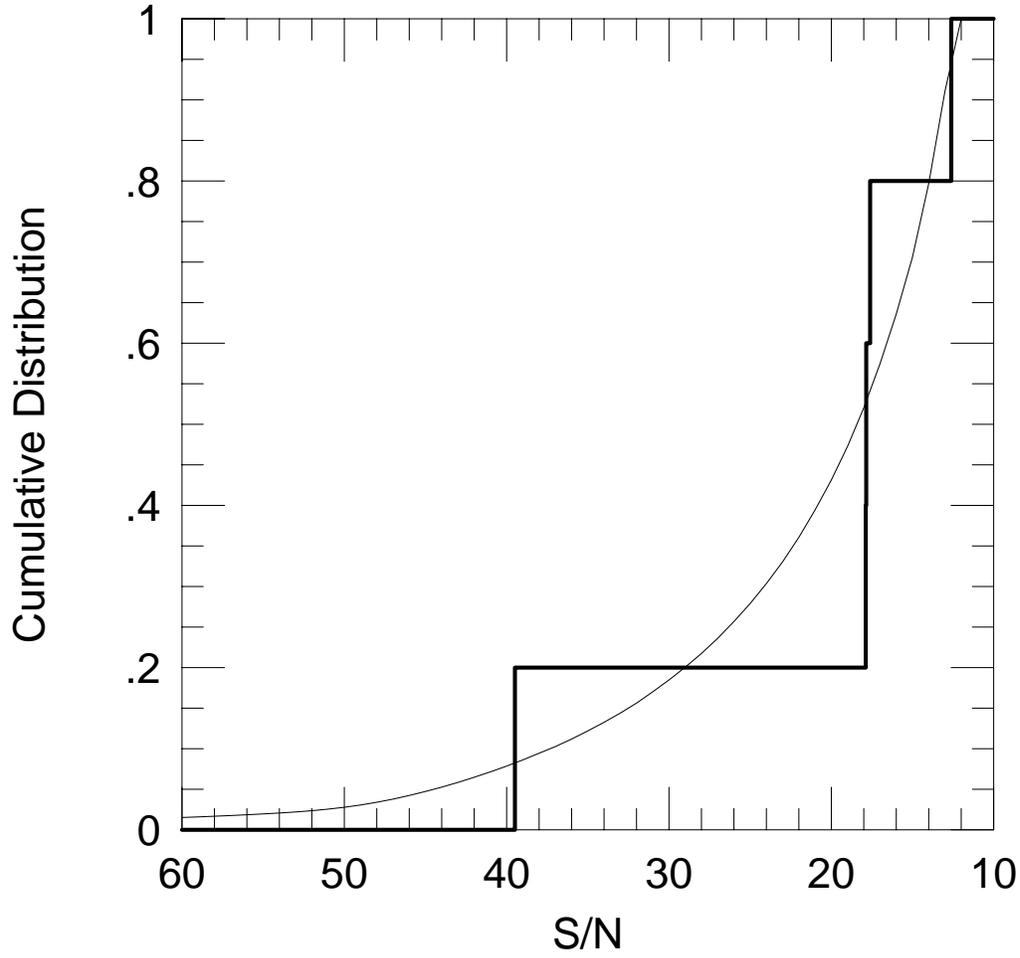}
\caption{\label{fig:kssn}
Cumulative distribution as a function of $S/N$
of the 5 confirmed OGLE transiting planets
({\it bold histogram}) compared to the distribution expected from
our model ({\it solid curve}). A KS test shows the the two are
consistent at $P(d=0.342)=0.50$. 
}
\end{figure}

\begin{figure}
\plotone{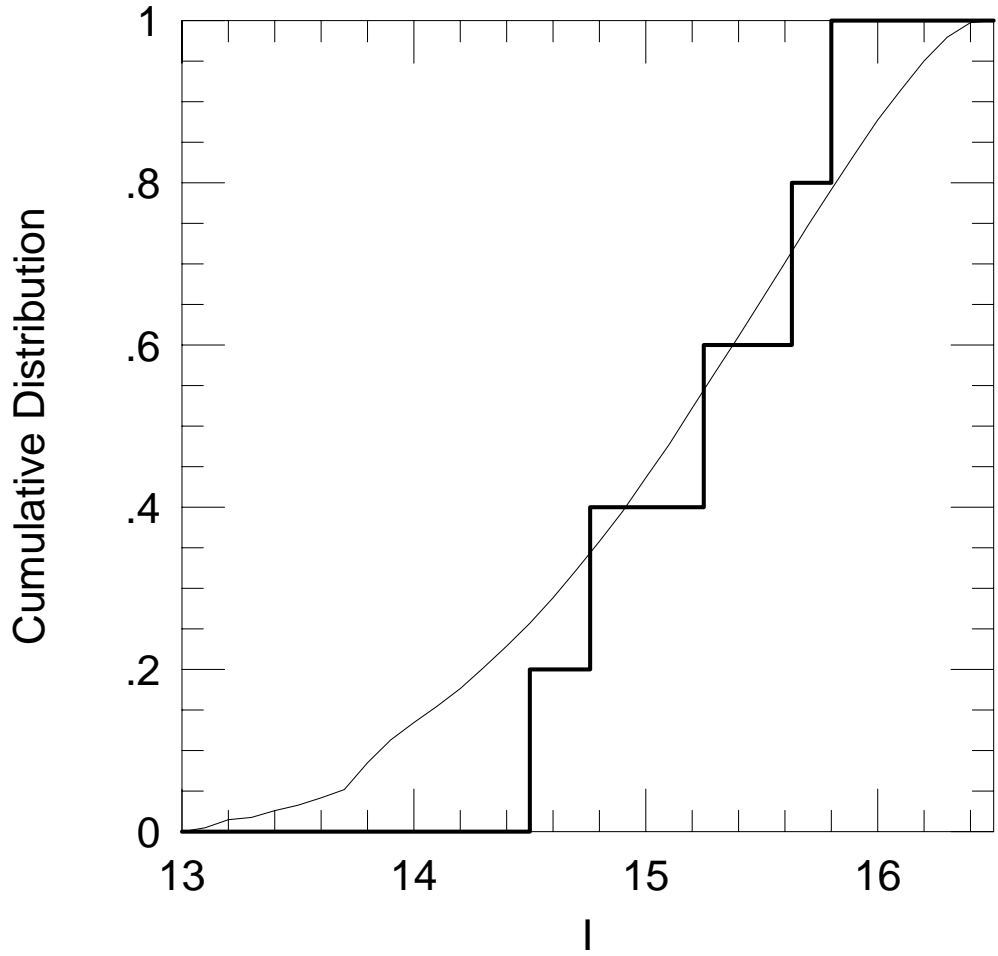}
\caption{\label{fig:ksi}
Cumulative distribution as a function of $I$-band apparent magnitude
of the 5 confirmed OGLE transiting planets
({\it bold histogram}) compared to the distribution expected from
our model ({\it solid curve}). A KS test shows the the two are
consistent at $P(d=0.257)=0.82$.
}
\end{figure}

\begin{figure}
\plotone{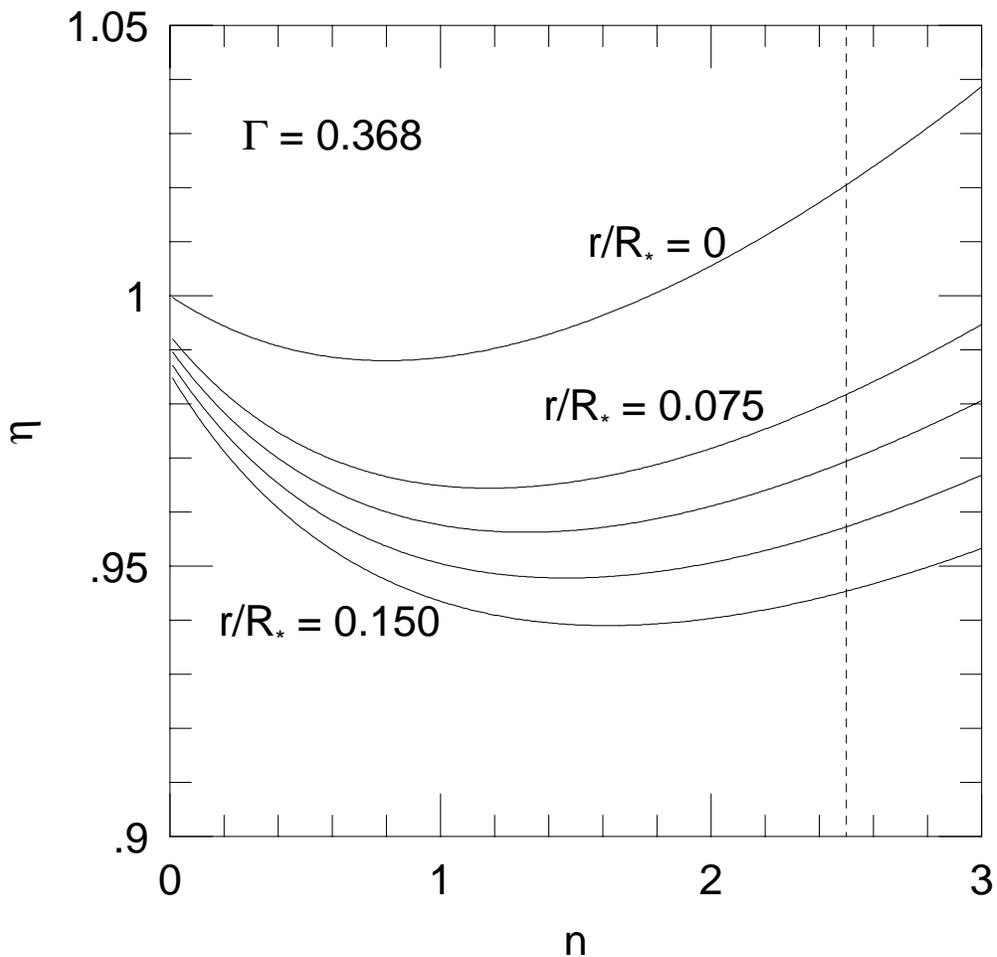}
\caption{\label{fig:ld}
Combined effect of limb darkening and ingress/egress on the number
of systems probed for planets relative to the case of a point-planet
occulting a star of uniform surface brightness.  The five curves
show the degradation factor $\eta$ as a function of power-law index $n$,
for planet-star $r/R_*=0$, 0.075, 0.1, 0.125 and 0.15.  The power-law
index parameterizes the growth of survey sensitivity as a function of
$S/N$ threshold, i.e. $\propto (S/N)_{\rm min}^{-n}$.  The dashed
line indicates $n=2.5$, the value that most closely represents
the behavior of the present survey.}
\end{figure}

\begin{figure}
\plotone{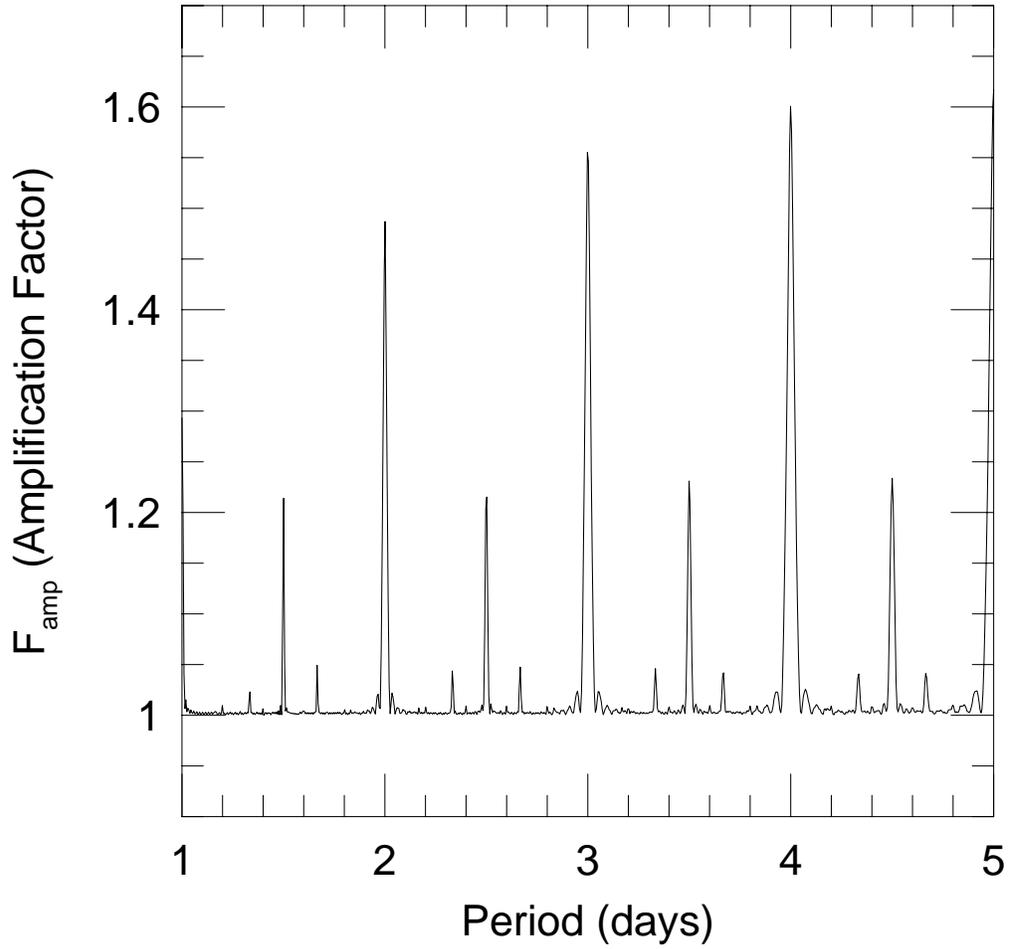}
\caption{\label{fig:A}
Period versus amplification factor $F_{amp}$ for a 24-hour day, 
6-hour night, 10-minute observations, an 80-day campaign, and a fraction 
$f=1/30(P/4\,{\rm days})^{-2/3}$ of the orbit spent in transit.
}\end{figure}

As illustrated in Figure~\ref{fig:8}, the OGLE-III target sample
is dominated by G stars, roughly 2/3 of which are expected
to have binary companions \citep{dm91}.  At their typical
distances of 1 kpc (see Fig.~\ref{fig:8}), most of these
binaries would be unresolved.  We ignore
this blending by binaries when modeling the search process.
However, we argue here that this has only a very small impact
on the implied sensitivity of the survey.

We first note that the underlying {\it Hipparcos} data, which
are used to model the field sample, are affected by exactly
the same phenomenon.  The physical separation at which
blending sets in for the {\it Hipparcos} sample is about
a decade smaller than for the field sample because the former is about
20 times closer, while the {\it Hipparcos} resolution is somewhat 
worse than that of OGLE.  However, since there are roughly equal
numbers of binaries per decade over 7 decades of separation, this
is a relatively small difference.  Hence the merged binaries are
automatically assigned the correct combined luminosity by our
modeling procedure.

For simplicity, let us now consider the special case in which
the two components of the binary have the same surface brightness.
Let the two stars have flux fractions $f$ and $(1-f)$,
and let us initially focus on a planet transiting
the first component.  Our model will assign a radius to the
star that is too large by a factor $f^{-1/2}$.  Consequently,
it will overestimate the duration of the transits by the same factor. 
This will in turn lead to an overestimate of the $S/N$ generated
by the transits by a factor $f^{-1/4}$.
At fixed semimajor axis, this error in the radius estimate 
will also cause us overestimate the probability of a transit by
a factor $f^{-1/2}$.  Of more direct interest is the error
at fixed period.  In this paper, we have considered two mass-radius
relations $M = R^m$, where $m=1$ for stars dimmer than the Sun
and $m=0$ for stars more luminous than the Sun.  Using Kepler's Third Law,
our overestimate of the transit probability is therefore
$f^{-1/2 + m/6}$.  Combining these factors, and assuming the
sensitivity of the survey grows $\propto (S/N)^{-n}$, we find that
the sensitivity of the binary to planets relative to our 
erroneous treatment of it as a single star is
\begin{equation}
F_{\rm bin}(f) = f^{n/4 + 1/2 - m/6} + (1- f)^{n/4 + 1/2 - m/6}.
\label{eqn:binary}
\end{equation}
This expression deviates maximally from unity at $f=1/2$
(i.e., equal-mass binaries), for
which $F_{\rm bin}(1/2) = 2^{1/2 + m/6 - n/4}$.  For $n=2.5$,
which is the appropriate approximation for the current survey,
we obtain $F_{\rm bin} = 0.92$ and 1.03 for $m=0$ and $m=1$,
respectively.  Since the majority of binaries are relatively
far from equal-mass \citep{dm91}, the typical effects are smaller.
Hence, ignoring source binarity has relatively little impact on
the results.

\section{Scaling Relations
\label{sec:scaling}}

\begin{figure}
\plotone{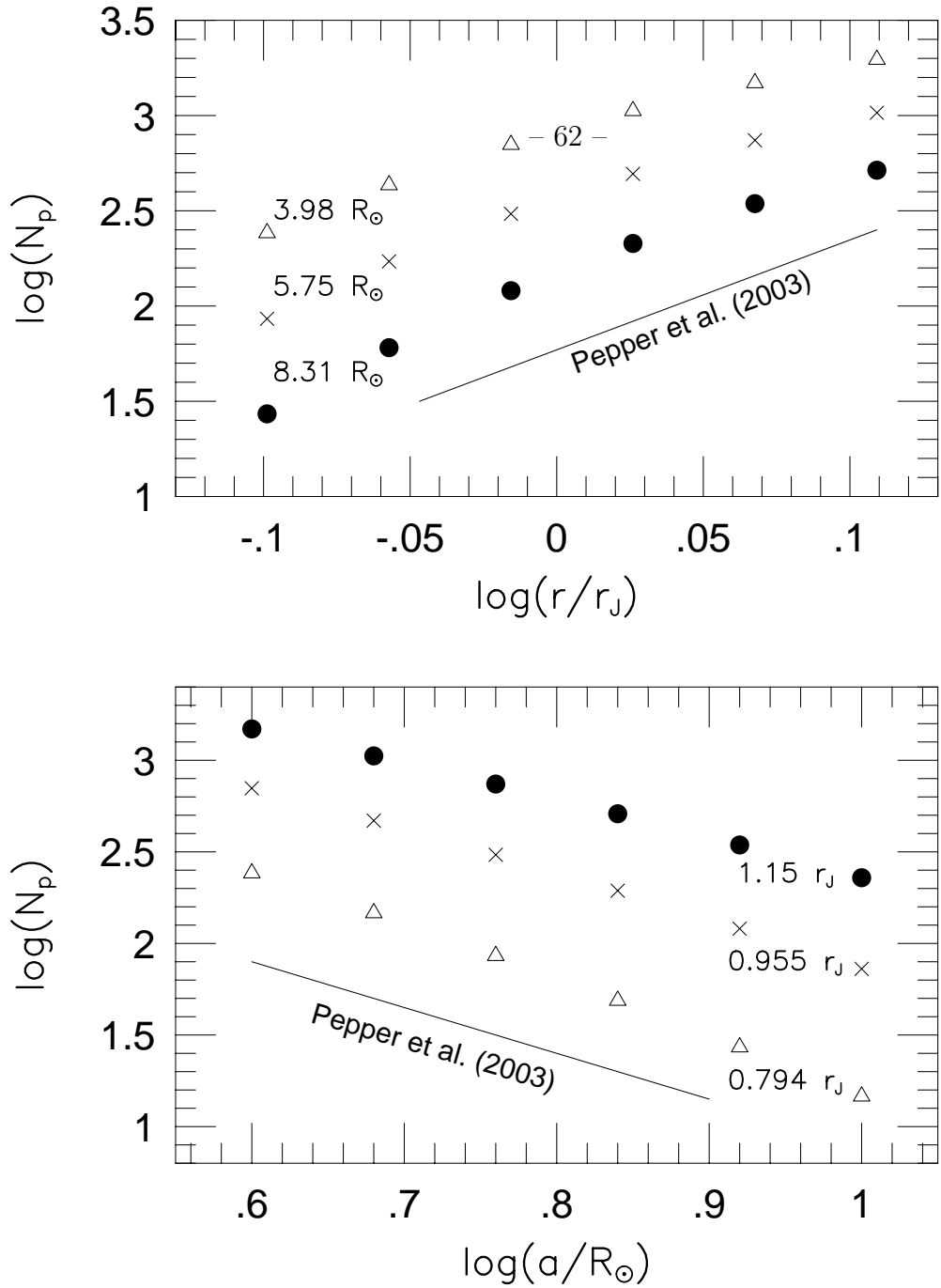}
\caption{\label{fig:scaling} The top panel shows the number of systems
probed for planets ($N_p$) in the OGLE-III survey toward Carina
as a function of the
planet radius, $r$, for three values of the orbital semi-major axis,
 $a=3.98\,R_\odot$ ({\it triangles}), $a=5.75\,R_\odot$ ({\it crosses}), and
$a=8.31\,R_\odot$ ({\it circles}).  
The lower panel shows $N_p$ as a function of $a$ for
three different values of the planet radius,
$r=0.794\, r_J$ ({\it triangles}), $r=0.955\, r_J$ ({\it crosses}), and
$r=1.15\, r_J$ ({\it circles}).
The solid lines in each panel represent the 
slopes of the scaling relations
predicted by \citet{scaling}, $N_p\propto a^{-5/2}$ and $N_p\propto r^6$,
respectively.
}
\end{figure}

\citet{scaling} predict the scaling relations between the number of
systems probed ($N_p$) and both the semi-major axis of the planet's
orbit ($a$) and the planet's radius ($r$), $N_p\propto
a^{-5/2}r^{6}$~\citep{scaling}. The top panel of
Figure~\ref{fig:scaling} shows the number probed in the Carina field
as a function of $r$ for several different values of $a$, and the
bottom panel shows $N_p$ in Carina as a function of $a$ for several
values of $r$. The best-fit slopes (or power law indices) of the $a$
scaling relation in the Carina field are shown in
Table~\ref{tab:aslopes}, and the best-fit slopes of the $r$
scaling relation in the Carina field are shown in
Table~\ref{tab:rslopes}. The results of our simulation are in
qualitative agreement with the predictions of \citet{scaling}, which
were derived under the idealized assumptions of
uniform density, photon-limited noise, and
no extinction.  However, note that the actual power-law indices can
deviate from these ideal values by up to 25\% over the ranges explored
in Tables~\ref{tab:aslopes} and \ref{tab:rslopes}, so caution is
warranted when applying these relations.

\section{Conclusions
\label{sec:conclusions}}

We have measured the frequency of HJs ($3\,{\rm days}< P < 5\,{\rm days}$)
and VHJs ($1\,{\rm day}< P < 3\,{\rm days}$) by calculating the numbers
of stars probed for planets in the OGLE-III Carina and Galactic bulge
transit fields, and comparing these to the 5 planets actually detected.
At 90\% confidence, we find that the HJ frequency is 
$(1/310)(1^{+1.39}_{-0.59})$
and the VHJ frequency is $(1/690)(1^{+1.10}_{-0.54})$.
The HJ rate is consistent with the results of RV studies,
and the VHJ rate is marginally consistent with the complete lack of 
RV detections for $P<2\,{\rm days}$.  
Because RV surveys
are heavily biased toward metal-rich stars, which are known to harbor
substantially more planets than average, we expect them to yield
a higher rate of planet detection.  However, the statistics are too sparse
yet to be sensitive to this anticipated effect.

We find that the number of stars probed (to the limit set by the
followup threshold of $V_{\rm max}<17.5$) is less than number of
stars in the field to this limit by roughly a factor 6 toward Carina
and a factor 25 toward the bulge.  The remainder of the stars have
physical radii that are too large to permit a detectable transit signal.
Hence, a careful determination of the number of {\it accessible} stars in
a given field is required to produce reliable estimates of the planet
frequency.

The five OGLE-III planets are consistent with being uniformly distributed
between 1.00 and 1.25 Jovian radii, although the number of detections is
too small to probe this distribution in detail.  
The survey has deteriorating sensitivity
at smaller radii, but would have detected planets
of larger radius, had a substantial such population been present.  We place
a 95\% conficence upper limit of $F<1/400$ on planets with $1.3<r/r_J<1.5$
in the period range of 1--5 days.

Additionally, we have investigated the impact of four effects
on the senstivity of the OGLE-III survey to planets with periods of
1--5 days and found this to be of order a few percent in each case.
These include limb darkening, ingress/egress, the periodic nature of 
ground-based observations, and unresolved binaries.  The first
two were specifically included in our simulations while the latter
two were not.

Finally, we have also shown that the scaling relations predicted by
\citet{scaling} are approximately valid for planets probed in the
OGLE-III survey, but that the actual power-law indices can deviate
from the predicted ones by up to 25\% over the parameter ranges
we explored.

\acknowledgments

We are grateful to Fr\'ed\'eric Pont, whose extensive comments, in 
several iterations, added greatly to our understanding of the
selection effects operating in the OGLE survey.
Work by AG and SD was supported by grant AST-0452758 from the
NSF.  
 Work by BSG was supported by a Menzel Fellowship from the Harvard College
Observatory.
A.U.\ was supported by the following grants: Polish MNII  2P03D02124,
``Stypendium Profesorskie'' of the Foundation for Polish Science, NSF
AST-0204908 and NASA NAG5-12212.
Any opinions, findings, and conclusions or 
recommendations expressed in this material are those of the authors
and do not necessarily reflect the views of the NSF.


\clearpage

\begin{deluxetable}{r l l l}
\tablecaption{\label{tab:vetting} Cuts in Data and Model}
\tablewidth{0pt} \tablehead{ \colhead{Number} & \colhead{Cut/Procedure} &
\colhead{Data} & \colhead{Model}
}
\startdata
 1. & Data Reduction & Yes & No \\
 2. & Systematics Removal & Yes & No \\
 3. & Saturation     & Yes & Yes \\
 4. & $\delta<0.08\,$mag  & Yes & Yes \\
 5. & $1.052<(P/{\rm day})<10$   & Yes & Yes \\
 6. & $\sigma<0.015\,$mag & Yes & Yes \\
 7. & $\alpha >9$    & Yes? & Yes \\
 8. & $SDE>{\rm max}[3,(4.9 - 0.1\alpha)]$ & Yes & Yes? \\
 9. & color-mag (bulge only) & Yes & Yes \\
10. & $>3$ (or 2) Transits  & Yes & Yes? \\
11. & $\delta<0.04\,$mag  & Yes & Yes \\
12. & $P<5\,$days          & Yes & Yes \\
13. & $\alpha>11$    & Yes? & Yes \\
14. & Period Recovery  & Yes & Yes? \\
15. & By-eye Culling & Yes & Yes? \\
16. & Eclipsing Binaries & Yes & No \\
17. & Rapid Rotators & Yes & No \\
18. & $(V-I)_0>0.4$  & Yes? & Yes \\
19. & $V<17.5$       & Yes? & Yes \\
20. & Unresolved Binaries & Yes & No \\
\enddata
\end{deluxetable}

\clearpage

\begin{deluxetable}{l l l l l r l l l r}
\tablecaption{\label{tab:planets} Confirmed Transiting Planets}
\tablewidth{0pt} \tablehead{ \colhead{Name} & \colhead{Field} &
\colhead{Period} & \colhead{$I$} & \colhead{$\Delta I$} &
\colhead{$N_{\rm tr}$} & \colhead{$r/r_J$} &
\colhead{$R_*/R_\odot$} &
\colhead{$M/M_J$} &
\colhead{$S/N$}
\\
\colhead{} &
\colhead{} &
\colhead{(days)} &
\colhead{} &
\colhead{} &
\colhead{} &
\colhead{} &
\colhead{} &
\colhead{} &
\colhead{}
}
\startdata 
OGLE-TR-10 & bulge & 3.1014 &14.76 & 0.019 & 7 & 1.16 & 1.18 & 0.54 &17.61\\ 
OGLE-TR-56 & bulge & 1.21193 &15.25 & 0.013 & 12 & 1.23 & 1.10 & 0.9 &17.88\\ 
OGLE-TR-111 & Carina & 4.01610 &15.63 & 0.019 & 9 & 1.00 & 0.85 & 0.53 &17.85\\ 
OGLE-TR-113 & Carina & 1.43250 &14.50 & 0.030 & 10 & 1.09 & 0.78 & 1.35&39.49\\ 
OGLE-TR-132 & Carina & 1.68965 &15.80 & 0.011 & 11 & 1.13 & 1.43 & 1.19 &12.61\\
\enddata
\tablecomments{The mass and
radius estimates for OGLE-TR-10 are from \citet{holman}, while those
of OGLE-TR-132 are from \citet{moutou04}.
The S/N is derived from the original data using the \citet{kovacs}
algorithm and eq.~(\ref{eqn:alphaeq}).  The remaining
data are from the papers in which the candidates were announced: 
\citep{bulge,alpha,carina,systematics}; or confirmed as planets
\citep{konacki03a,konacki-tr-10,tr-113-132,tr-111}.}
\end{deluxetable}

\clearpage

\begin{deluxetable}{l l l l l r}
\tablecaption{\label{tab:unknown} Unclassified Candidates}
\tablewidth{0pt}
\tablehead{
\colhead{Name} &
\colhead{Field} &
\colhead{Period} &
\colhead{$I$} &
\colhead{$\Delta I$} &
\colhead{$N_{\rm tr}$}
\\
\colhead{} &
\colhead{} &
\colhead{(days)} &
\colhead{} &
\colhead{} &
\colhead{}
}
\startdata
OGLE-TR-51 & bulge & 1.74812 & 16.71 & 0.034 & 5\\
OGLE-TR-134 & bulge & 4.53720 & 13.49 & 0.011 & 2\\
OGLE-TR-137 & bulge & 2.53782 & 15.85 & 0.030 & 7\\
OGLE-TR-82 & Carina &  0.76416 & 16.30 & 0.034 & 22
\enddata
\end{deluxetable}

\clearpage

\begin{deluxetable}{l l r l l r l}
\tablecaption{\label{tab:goodstars} Fraction of Stars Probed}
\tablewidth{0pt} 
\tablehead{ 
&\multicolumn{3}{c}{Carina} &\multicolumn{3}{c}{Bulge}\\
\colhead{$V_{\rm max}$} &
\colhead{$N_S$} &
\colhead{$N_D$} &
\colhead{$N_S/N_D$} &
\colhead{$N_S$} &
\colhead{$N_D$} &
\colhead{$N_S/N_D$}}
\startdata
15.5 &  930 &  8490 & 0.11 & 1057 &   7641 & 0.138 \\ 
16.0 & 1766 & 12816 & 0.14 & 1634 &  13057 & 0.125 \\ 
16.5 & 2969 & 18510 & 0.16 & 2309 &  23517 & 0.098 \\ 
17.0 & 4486 & 27344 & 0.16 & 2909 &  42930 & 0.068 \\ 
17.5 & 6153 & 39207 & 0.16 & 3287 &  79709 & 0.041 \\ 
18.0 & 7542 & 54430 & 0.14 & 3397 & 156326 & 0.022 \\ 
18.5 & 8382 & 74506 & 0.11 & 3414 & 278524 & 0.012 \\ 
\enddata
\tablecomments{$N_S$ is the number of stars around which a planet 
orbiting edge on with $r=1.2\, r_J$ and $a=7.94\, R_\odot$ could be detected,
while $N_D$ is the total number of stars in the field.}
\end{deluxetable}


\begin{deluxetable}{r r r r r r r r}
\tablecaption{\label{tab:byperiod} Planet Frequency by Period}
\tablewidth{0pt} 
\tablehead{ 
&\multicolumn{2}{c}{Carina} &\multicolumn{2}{c}{Bulge} &
\multicolumn{2}{c}{Total}&\\
\colhead{$P$/day}& 
\colhead{$N_p$} &
\colhead{Planets} &
\colhead{$N_p$} & 
\colhead{Planets} &
\colhead{$N_p$} &
\colhead{Planets} &
\colhead{Frequency}
}
\startdata 
1.0-2.0 &  885 & 2 &  495 & 1 & 1380 & 3 & 0.22\% \\
2.0-3.0 &  456 & 0 &  226 & 0 &  682 & 0 & \\
3.0-4.0 &  266 & 0 &  120 & 1 &  386 & 1 & 0.26\% \\
4.0-5.0 &  163 & 1 &   69 & 0 &  232 & 1 & 0.43\% \\
\enddata
\tablecomments{
Values are averaged over planet radii ranging from $1.00\,
r_J$ to $1.25\, r_J$. 
}
\end{deluxetable}

\clearpage

\begin{deluxetable}{r r r r r r r r}
\tablecaption{\label{tab:byradius} Planet Frequency by Radius}
\tablewidth{0pt} 
\tablehead{ 
&\multicolumn{2}{l}{Carina} &\multicolumn{2}{l}{Bulge} &\multicolumn{2}{l}{Total}&\\
\colhead{$r/r_J$}& 
\colhead{$N_p$} &
\colhead{Planets} & 
\colhead{$N_p$} &
\colhead{Planets} & 
\colhead{$N_p$} &
\colhead{Planets} & 
\colhead{Frequency}}
\startdata 
0.60 &    5 & 0 &    1 & 0 &    6 & 0 & \\
0.65 &   10 & 0 &    4 & 0 &   14 & 0 & \\
0.70 &   20 & 0 &    8 & 0 &   28 & 0 & \\
0.75 &   35 & 0 &   15 & 0 &   50 & 0 & \\
0.80 &   58 & 0 &   25 & 0 &   83 & 0 & \\
0.85 &   91 & 0 &   39 & 0 &  130 & 0 & \\
0.90 &  132 & 0 &   58 & 0 &  190 & 0 & \\
0.95 &  184 & 0 &   83 & 0 &  267 & 0 & \\
1.00 &  244 & 1 &  113 & 0 &  357 & 1 & 0.28\% \\
1.05 &  314 & 0 &  150 & 0 &  464 & 0 & \\
1.10 &  392 & 1 &  194 & 0 &  586 & 1 & 0.17\% \\
1.15 &  477 & 1 &  244 & 1 &  721 & 2 & 0.28\% \\
1.20 &  567 & 0 &  301 & 0 &  868 & 0 & \\
1.25 &  661 & 0 &  363 & 1 & 1024 & 1 & 0.10\% \\
1.30 &  759 & 0 &  432 & 0 & 1191 & 0 & \\
1.35 &  856 & 0 &  503 & 0 & 1359 & 0 & \\
1.40 &  952 & 0 &  577 & 0 & 1529 & 0 & \\
1.45 & 1046 & 0 &  654 & 0 & 1700 & 0 & \\
1.50 & 1136 & 0 &  731 & 0 & 1867 & 0 & \\
1.55 & 1225 & 0 &  811 & 0 & 2036 & 0 & \\
1.60 & 1312 & 0 &  893 & 0 & 2205 & 0 & \\
1.65 & 1393 & 0 &  974 & 0 & 2367 & 0 & \\
1.70 & 1467 & 0 & 1052 & 0 & 2519 & 0 & \\
\enddata
\tablecomments{
Values are averaged over periods from one to five days. See
Fig.~\ref{fig:logbyradius} for averages broken into two intervals:
$1<P/{\rm day}<3$ and $3<P/{\rm day}<5$.
}
\end{deluxetable}

\clearpage

\begin{deluxetable}{l r r r r}
\tablecaption{\label{tab:peakfrac} Fraction in Resonances}
\tablewidth{0pt}
\tablehead{
\colhead{In Resonance} &
\colhead{2 days} &
\colhead{3 days} &
\colhead{4 days} &
\colhead{Total}}
\startdata
Percent of Periods     & 1.3\% & 1.9\% & 2.6\% & 5.8\%\\
Percent of Sensitivity & 1.6\% & 2.5\% & 3.3\% & 7.4\%
\enddata
\end{deluxetable}

\clearpage

\begin{deluxetable}{r l}
\tablecaption{\label{tab:aslopes} Semimajor-Axis Power Laws}
\tablewidth{0pt}
\tablehead{
\colhead{$\log(r/r_J)$} &
\colhead{$\gamma$ ($N_p=a^\gamma$)}}
\startdata
$-0.10$ & $-3.05\pm 0.05$\\
$-0.06$ & $-2.75\pm 0.06$\\
$-0.02$ & $-2.46\pm 0.05$\\
$ 0.02$ & $-2.23\pm 0.04$\\
$ 0.06$ & $-2.03\pm 0.04$\\
$ 0.10$ & $-1.86\pm 0.03$\\
\enddata
\tablecomments{\citet{scaling} predict $N_p\propto a^\gamma$ with 
$\gamma=-5/2$ for transit surveys
of nearby stars}
\end{deluxetable}


\begin{deluxetable}{r l}
\tablecaption{\label{tab:rslopes} Planet-Radius Power Laws}
\tablewidth{0pt}
\tablehead{
\colhead{$\log(a/R_\odot)$} &
\colhead{$\gamma$ ($N_p=r^\gamma$)}}
\startdata
0.60 & $4.5\pm 0.3$\\
0.68 & $4.9\pm 0.3$\\
0.76 & $5.4\pm 0.4$\\
0.84 & $5.9\pm 0.4$\\
0.92 & $6.4\pm 0.4$\\
1.00 & $6.9\pm 0.4$\\
\enddata
\tablecomments{\citet{scaling} predict $N_p\propto r^\gamma$ with
$\gamma=6$ for transit surveys of nearby stars.}
\end{deluxetable}

\clearpage

\end{document}